\documentclass[revtex4]{emulateapj}
\usepackage{natbib}
\usepackage{comment}
\usepackage{lscape}
\usepackage{amsmath}
\usepackage{hyperref,breakurl}

\shorttitle{GBT 33\,GHz OBSERVATIONS OF GALAXY NUCLEI AND EXTRA-NUCLEAR STAR-FORMING REGIONS}
\shortauthors{MURPHY ET AL.}

\shorttitle{SFRS: VLA Ka-BAND OBSERVATIONS}
\shortauthors{MURPHY ET AL.}
\slugcomment{Submitted: August 7, 2017; Accepted: November 8, 2017}

\begin{document}
\title{The Star Formation in Radio Survey: Jansky Very Large Array 33\,GHz Observations of Nearby Galaxy Nuclei and Extranuclear Star-Forming Regions}

\author{E.J.\,Murphy\altaffilmark{1,2}, D.\,Dong\altaffilmark{3,4}, E.\,Momjian\altaffilmark{5}, S.\,Linden\altaffilmark{6 }, R.C.~Kennicutt,~Jr.\altaffilmark{7} D.S.\,Meier\altaffilmark{8,5}, E.\,Schinnerer\altaffilmark{9}, and J.L.\,Turner\altaffilmark{10}}

%, J.J.\,Condon\altaffilmark{5}, E.\,Schinnerer\altaffilmark{6}, L.\,Armus \altaffilmark{7}, G.\,Helou\altaffilmark{8}, J.L.\,Turner\altaffilmark{9}}
\altaffiltext{1}{National Radio Astronomy Observatory, 520 Edgemont Road, Charlottesville, VA 22903, USA; emurphy@nrao.edu}
\altaffiltext{2}{Infrared Processing and Analysis Center, California Institute of Technology, MC 220-6, Pasadena CA, 91125, USA}
%\altaffiltext{3}{Observatories of the Carnegie Institution for Science, 813 Santa Barbara Street, Pasadena, CA 91101, USA}
\altaffiltext{3}{Department of Physics and Astronomy, Pomona College, Claremont, CA 91711, USA}
\altaffiltext{4}{California Institute of Technology, MC 100-22, Pasadena, CA 91125, USA}
\altaffiltext{5}{National Radio Astronomy Observatory, P.O. Box O, 1003 Lopezville Road, Socorro, NM 87801, USA}
\altaffiltext{6}{Department of Astronomy, University of Virginia, 3530 McCormick Road,Charlottesville, VA 22904, USA}
\altaffiltext{7}{Institute of Astronomy, University of Cambridge, Madingley Road, Cambridge CB3 0HA, UK}
\altaffiltext{8}{New Mexico Institute of Mining and Technology, 801 Leroy Place, Socorro, NM 87801, USA}
\altaffiltext{9}{Max Planck Institut f\"{u}r Astronomie, K\"{o}nigstuhl 17, Heidelberg D-69117, Germany}
\altaffiltext{10}{Department of Physics and Astronomy, UCLA, Los Angeles, CA 90095, USA}

%\altaffiltext{5}{National Radio Astronomy Observatory, 520 Edgemont Road, Charlottesville, VA 22903, USA}
%\altaffiltext{6}{Max Planck Institut f\"{u}r Astronomie, K\"{o}nigstuhl 17, Heidelberg D-69117, Germany}
%\altaffiltext{7}{ {\it Spitzer Science Center,} California Institute of Technology, MC 314-6, Pasadena CA, USA}
%\altaffiltext{8}{California Institute of Technology, MC 100-22, Pasadena, CA 91125, USA}
%\altaffiltext{9}{Department of Physics and Astronomy, UCLA, Los Angeles, CA 90095, USA}

\begin{abstract}
We present 33\,GHz imaging for 112~pointings towards galaxy nuclei and extranuclear star-forming regions at $\approx$2\arcsec~resolution using the Karl~G.~Jansky Very~Large~Array (VLA) as part of the Star Formation in Radio Survey.    
A comparison with 33\,GHz Robert C.~Byrd~Green~Bank~Telescope single-dish observations indicates that the interferometric VLA observations recover $78\pm4\%$ of the total flux density over 25\arcsec~regions~($\approx$\,kpc-scales) among all fields.  
On these scales, the emission being resolved out is most likely diffuse non-thermal synchrotron emission.  
Consequently, on the $\approx30-300$\,pc scales sampled by our VLA observations, the bulk of the 33\,GHz emission is recovered and primarily powered by free-free emission from discrete H{\sc ii} regions, making it an excellent tracer of massive star formation.  
Of the 225 discrete regions used for aperture photometry, 162 are extranuclear (i.e., having galactocentric radii $r_{\rm G} \geq 250$\,pc) and detected at $>3\sigma$ significance at 33\,GHz and in H$\alpha$.  
Assuming a typical 33\,GHz thermal fraction of 90\%, the ratio of optically-thin 33\,GHz-to-uncorrected H$\alpha$ star formation rates indicate a median extinction value on $\approx30-300$\,pc scales of $A_{\rm H\alpha} \approx 1.26\pm0.09$\,mag with an associated median absolute deviation of 0.87\,mag.  
We find that 10\% of these sources are ``highly embedded" (i.e., $A_{\rm H\alpha}\gtrsim3.3$\,mag), suggesting that on average H{\sc ii} regions remain embedded for $\lesssim1$\,Myr. 
Finally, we find the median 33\,GHz continuum-to-H$\alpha$ line flux ratio to be statistically larger within  $r_{\rm G}<250$\,pc relative the outer-disk regions by a factor of $1.82\pm0.39$, while the ratio of 33\,GHz-to-24\,$\mu$m flux densities are lower by a factor of $0.45\pm0.08$, which may suggest increased extinction in the central regions.
\end{abstract}

\section{Introduction}
%Radio emission from galaxies, nominally spanning $1-100\,$GHz, is powered by a variety of physical processes.  
%Although it is energetically weak with respect to a galaxy's bolometric luminosity, 
Radio emission from galaxies is powered by a combination of distinct physical processes.  
And although it is energetically weak with respect to a galaxy's bolometric luminosity, it provides critical information on the massive star formation activity, as well as access to the relativistic [magnetic field $+$ cosmic rays (CRs)] component in the interstellar medium (ISM) of galaxies.  
%insight into a number of physical processes shaping their evolution.    
%The most important of these is arguably massive star formation, which is why radio observations make excellent diagnostics for the current star formation rate (SFR) of galaxies.  
%These processes include massive star formation and associated supernova explosions, cosmic-ray (CR) electron and magnetic field distributions.  

Stars more massive than $\sim8\,M_{\sun}$ end their lives as core-collapse supernovae, whose remnants are thought to be the primary accelerators of CR electrons \citep[e.g.,][]{kk95} giving rise to the diffuse synchrotron emission observed from star-forming galaxies \citep{jc92}.  
These same massive stars are also responsible for the creation of H{\sc ii} regions that produce radio free-free emission, whose strength is directly proportional to the production rate of ionizing (Lyman continuum) photons. 

Radio frequencies spanning $\sim$$1-100$\,GHz, which are observable from the ground, are particularly useful in probing such processes.  
The non-thermal emission component typically has a steep spectrum ($S_{\nu} \propto \nu^{-\alpha}$, where $\alpha \sim 0.8$), while the thermal (free-free) component is relatively flat \citep[$\alpha\sim0.1$; e.g., ][]{jc92}.  
Accordingly, for globally integrated measurements of star-forming galaxies, lower frequencies (e.g., 1.4\,GHz) are generally dominated by non-thermal emission, while the observed thermal fraction of the emission increases with frequency, eventually being dominated by free-free emission once beyond $\sim$30\,GHz \citep{cy90}.  
For typical H{\sc ii} regions, the thermal fraction at 33\,GHz can be considerably higher, being $\sim$80\% \citep{ejm11b}.   
Thus, observations at such frequencies, which are largely unbiased by dust, provide an excellent diagnostic for the current star formation rate (SFR) of galaxies.  

%While free-free is appears to dominate the 33\,GHz emission from galaxies, it is worth pointing out that this may not always be the case.  
It is worth noting that the presence of an anomalous microwave emission (AME) component in excess of free-free emission between $\sim$10 and 90\,GHz, generally attributed to electric dipole rotational emission from ultrasmall ($a \la 10^{-6}$\,cm) grains \citep[e.g.,][]{wce57,dl98a,dl98b, plsd11} or magnetic dipole emission from thermal fluctuations in the magnetization of interstellar dust grains \citep[][]{dl99, bh16}, may complicate this picture.
%While generally attributed to rapidly rotating ultrasmall ($a \la 10^{-6}$\,cm) grains with a non-zero electric moment \citep[e.g.,][]{dl98b}, 
For a single outer-disk star-forming region in NGC\,6946, \citet{ejm10} reported an excess of 33\,GHz emission relative to what is expected given existing lower frequency radio data. 
This result has been interpreted as the first detection of so-called ``anomalous" dust emission outside of the Milky Way. 
While the excess was only detected for a single region in this initial study, follow-up observations yielded additional detections in the disk of NGC\,6946 \citep{bh14}.  
However it appears that this emission component is most likely sub-dominant for globally integrated measurements.  
%Given that the excess was only detected for a single region, this emission component may be negligible for globally integrated measurements.  

\begin{deluxetable*}{l|cccccccc}
\tablecaption{Galaxy Properties and Nuclear Source Positions \label{tbl-1}}
\tabletypesize{\scriptsize}
\tablewidth{0pt}
\tablehead{
\colhead{Galaxy}  & \colhead{R.A.} & \colhead{Decl.} & \colhead{Type\tablenotemark{a}} & \colhead{Dist.\tablenotemark{b}} & \colhead{Nuc. Type\tablenotemark{c}}  & \colhead{$D_{25}$\tablenotemark{a}} & \colhead{$i$} & \colhead{P.A.\tablenotemark{a}} \\
\colhead{} & \colhead{(J2000)} & \colhead{(J2000)} & \colhead{} & \colhead{(Mpc)} & \colhead{} & \colhead{(arcmin)} & \colhead{($\degr$)} & \colhead{($\degr$)} 
}
   NGC\,0337   &$  0 0   ~59   ~50.3   $&$-  0 7   ~34   ~44$  &      SBd  &19.3  &                SF  &  $2.9 \times 1.8$  & 52  &               130\\
   NGC\,0628   &$  0 1   ~36   ~41.7   $&$+   15   ~46   ~59$  &      SAc  & 7.2  &           \nodata  & $10.5 \times 9.5$  & 25  &                25\\
   NGC\,0855   &$  0 2   ~14  ~0 3.7   $&$+   27   ~52   ~38$  &        E  &9.73  &                SF  &  $2.6 \times 1.0$  & 70  &   67\rlap{$^{d}$}\\
   NGC\,0925   &$  0 2   ~27   ~17.0   $&$+   33   ~34   ~43$  &     SABd  &9.12  &                SF  & $10.5 \times 5.9$  & 57  &               102\\
   NGC\,1097   &$  0 2   ~46   ~19.1   $&$-   30   ~16   ~28$  &      SBb  &14.2  &               AGN  &  $9.3 \times 6.3$  & 48  &               130\\
   NGC\,1266   &$  0 3   ~16  ~0 0.8   $&$-  0 2   ~25   ~38$  &      SB0  &30.6  &               AGN  &  $1.5 \times 1.0$  & 49  &  108\rlap{$^{d}$}\\
   NGC\,1377   &$  0 3   ~36   ~38.9   $&$-   20   ~54  ~0 6$  &       S0  &24.6  &           \nodata  &  $1.8 \times 0.9$  & 61  &                92\\
    IC\,0342   &$  0 3   ~46   ~48.5   $&$+   68  ~0 5   ~46$  &    SABcd  &3.28  &             SF(*)  & $21.4 \times 20.$  & 21  &  153\rlap{$^{d}$}\\
   NGC\,1482   &$  0 3   ~54   ~39.5   $&$-   20   ~30  ~0 7$  &      SA0  &22.6  &                SF  &  $2.5 \times 1.4$  & 57  &               103\\
   NGC\,2146   &$  0 6   ~18   ~37.7   $&$+   78   ~21   ~25$  &     Sbab  &17.2  &             SF(*)  &  $6.0 \times 3.4$  & 56  &                57\\
   NGC\,2403   &$  0 7   ~36   ~50.0   $&$+   65   ~36  ~0 4$  &    SABcd  &3.22  &             SF(*)  &$21.9 \times 12.3$  & 57  &               128\\
Holmberg\,II   &$  0 8   ~19   ~13.3   $&$+   70   ~43  ~0 8$  &       Im  &3.05  &           \nodata  &  $7.9 \times 6.3$  & 37  &                16\\
   NGC\,2798   &$  0 9   ~17   ~22.8   $&$+   41   ~59   ~58$  &      SBa  &25.8  &            SF/AGN  &  $2.6 \times 1.0$  & 70  &               160\\
   NGC\,2841   &$  0 9   ~22  ~0 2.7   $&$+   50   ~58   ~36$  &      SAb  &14.1  &               AGN  &  $8.1 \times 3.5$  & 66  &               147\\
   NGC\,2976   &$  0 9   ~47   ~15.3   $&$+   67   ~55  ~0 0$  &      SAc  &3.55  &                SF  &  $5.9 \times 2.7$  & 64  &               143\\
   NGC\,3049   &$  0 9   ~54   ~49.6   $&$+  0 9   ~16   ~17$  &     SBab  &19.2  &                SF  &  $2.2 \times 1.4$  & 51  &                25\\
   NGC\,3077   &$   10  ~0 3   ~19.1   $&$+   68   ~44  ~0 2$  &    I0pec  &3.83  &             SF(*)  &  $5.4 \times 4.5$  & 34  &                45\\
   NGC\,3190   &$   10   ~18  ~0 5.6   $&$+   21   ~49   ~55$  &     SAap  &19.3  &            AGN(*)  &  $4.4 \times 1.5$  & 73  &               125\\
   NGC\,3184   &$   10   ~18   ~16.7   $&$+   41   ~25   ~27$  &    SABcd  &11.7  &                SF  &  $7.4 \times 6.9$  & 21  &               135\\
   NGC\,3198   &$   10   ~19   ~54.9   $&$+   45   ~32   ~59$  &      SBc  &14.1  &                SF  &  $8.5 \times 3.3$  & 68  &                35\\
    IC\,2574   &$   10   ~28   ~48.4   $&$+   68   ~28  ~0 2$  &     SABm  &3.79  &             SF(*)  & $13.2 \times 5.4$  & 67  &                50\\
   NGC\,3265   &$   10   ~31  ~0 6.7   $&$+   28   ~47   ~48$  &        E  &19.6  &                SF  &  $1.3 \times 1.0$  & 39  &                73\\
   NGC\,3351   &$   10   ~43   ~57.8   $&$+   11   ~42   ~14$  &      SBb  &9.33  &                SF  &  $7.4 \times 5.0$  & 48  &                13\\
   NGC\,3521   &$   11  ~0 5   ~48.9   $&$-  0 0  ~0 2  ~0 6$  &    SABbc  &11.2  &         SF/AGN(*)  & $11.0 \times 5.1$  & 63  &               163\\
   NGC\,3621   &$   11   ~18   ~16.0   $&$-   32   ~48   ~42$  &      SAd  &6.55  &               AGN  & $12.3 \times 7.1$  & 55  &               159\\
   NGC\,3627   &$   11   ~20   ~15.0   $&$+   12   ~59   ~30$  &     SABb  &9.38  &               AGN  &  $9.1 \times 4.2$  & 64  &               173\\
   NGC\,3773   &$   11   ~38   ~13.0   $&$+   12  ~0 6   ~45$  &      SA0  &12.4  &                SF  &  $1.2 \times 1.0$  & 33  &               165\\
   NGC\,3938   &$   11   ~52   ~49.5   $&$+   44  ~0 7   ~14$  &      SAc  &17.9  &             SF(*)  &  $5.4 \times 4.9$  & 25  &   29\rlap{$^{d}$}\\
   NGC\,4254   &$   12   ~18   ~49.4   $&$+   14   ~24   ~59$  &      SAc  &14.4  &            SF/AGN  &  $5.4 \times 4.7$  & 30  &   24\rlap{$^{d}$}\\
   NGC\,4321   &$   12   ~22   ~54.9   $&$+   15   ~49   ~21$  &    SABbc  &14.3  &               AGN  &  $7.4 \times 6.3$  & 32  &                30\\
   NGC\,4536   &$   12   ~34   ~27.1   $&$+  0 2   ~11   ~17$  &    SABbc  &14.5  &            SF/AGN  &  $7.6 \times 3.2$  & 66  &               130\\
   NGC\,4559   &$   12   ~35   ~57.7   $&$+   27   ~57   ~36$  &    SABcd  &6.98  &                SF  & $10.7 \times 4.4$  & 67  &               150\\
   NGC\,4569   &$   12   ~36   ~49.8   $&$+   13  ~0 9   ~46$  &    SABab  &9.86  &               AGN  &  $9.5 \times 4.4$  & 64  &                23\\
   NGC\,4579   &$   12   ~37   ~43.6   $&$+   11   ~49  ~0 2$  &     SABb  &16.4  &               AGN  &  $5.9 \times 4.7$  & 37  &                95\\
   NGC\,4594   &$   12   ~39   ~59.4   $&$-   11   ~37   ~23$  &      SAa  &9.08  &               AGN  &  $8.7 \times 3.5$  & 69  &                90\\
   NGC\,4625   &$   12   ~41   ~52.4   $&$+   41   ~16   ~24$  &    SABmp  & 9.3  &                SF  &  $2.2 \times 1.9$  & 31  &   28\rlap{$^{d}$}\\
   NGC\,4631   &$   12   ~42  ~0 5.9   $&$+   32   ~32   ~22$  &      SBd  &7.62  &             SF(*)  & $15.5 \times 2.7$  & 83  &                86\\
   NGC\,4725   &$   12   ~50   ~26.6   $&$+   25   ~30  ~0 6$  &    SABab  &11.9  &               AGN  & $10.7 \times 7.6$  & 45  &                35\\
   NGC\,4736   &$   12   ~50   ~53.0   $&$+   41  ~0 7   ~14$  &     SAab  &4.66  &            AGN(*)  & $11.2 \times 9.1$  & 35  &               105\\
   NGC\,4826   &$   12   ~56   ~43.9   $&$+   21   ~41  ~0 0$  &     SAab  &5.27  &               AGN  & $10.0 \times 5.4$  & 59  &               115\\
   NGC\,5055   &$   13   ~15   ~49.2   $&$+   42  ~0 1   ~49$  &     SAbc  &7.94  &               AGN  & $12.6 \times 7.2$  & 56  &               105\\
   NGC\,5194   &$   13   ~29   ~52.7   $&$+   47   ~11   ~43$  &   SABbcp  &7.62  &               AGN  & $11.2 \times 6.9$  & 53  &               163\\
   NGC\,5398   &$   14  ~0 1   ~20.2   $&$-   33  ~0 4  ~0 9$  &     SBdm  &7.66  &           \nodata  &  $2.8 \times 1.7$  & 53  &               172\\
   NGC\,5457   &$   14  ~0 3   ~12.6   $&$+   54   ~20   ~57$  &    SABcd  & 6.7  &             SF(*)  & $28.8 \times 26.$  & 26  &   29\rlap{$^{d}$}\\
   NGC\,5474   &$   14  ~0 5  ~0 1.3   $&$+   53   ~39   ~44$  &     SAcd  & 6.8  &             SF(*)  &  $4.8 \times 4.3$  & 27  &   98\rlap{$^{d}$}\\
   NGC\,5713   &$   14   ~40   ~11.3   $&$-  0 0   ~17   ~27$  &   SABbcp  &21.4  &                SF  &  $2.8 \times 2.5$  & 27  &                10\\
   NGC\,5866   &$   15  ~0 6   ~29.5   $&$+   55   ~45   ~48$  &       S0  &15.3  &               AGN  &  $4.7 \times 1.9$  & 69  &               128\\
   NGC\,6946   &$   20   ~34   ~52.3   $&$+   60  ~0 9   ~14$  &    SABcd  & 6.8  &                SF  & $11.5 \times 9.8$  & 32  &   53\rlap{$^{d}$}\\
   NGC\,7331   &$   22   ~37  ~0 4.1   $&$+   34   ~24   ~56$  &      SAb  &14.5  &               AGN  & $10.5 \times 3.7$  & 72  &               171\\
   NGC\,7793   &$   23   ~57   ~49.2   $&$-   32   ~35   ~24$  &      SAd  &3.91  &                SF  &  $9.3 \times 6.3$  & 48  &                98  
\enddata
\tablenotetext{a}{Morphological types, diameters, and position angles were taken from the Third Reference Catalog of Bright Galaxies \citep[RC3;][]{rc3}.}%the NASA/IPAC Extragalactic Database (NED; http://nedwww.ipac.caltech.edu).}
\tablenotetext{b}{Redshift-independent distance taken from the list compiled by \citet{kf11}, except for the two non-KINGFISH galaxies NGC\,5194 \citep{rc02} and NGC\,2403 \citep{hkp01}.}
\tablenotetext{c}{Nuclear type based on optical spectroscopy: SF$\,=\,$ Star-Forming; AGN$\,=\,$ Non-thermal emission as given in Table~5 of \citet{jm10} or (*) Table~4 of \citet{hfs97}.}
\tablenotetext{d}{Position angle taken from \citet{2massLGA}.}
\end{deluxetable*}

Due to the faintness of galaxies at high (i.e., $\ga$15\,GHz) radio frequencies, existing work has been restricted to the brightest objects, and small sample sizes.  
For example, past studies demonstrating the link between high-frequency free-free emission and massive star formation include investigations of Galactic star-forming regions \citep[e.g.,][]{pgm67a}, nearby dwarf irregular galaxies \citep[e.g.,][]{kg86}, galaxy nuclei \citep[e.g.,][]{th83,th94}, nearby starbursts \citep[e.g.,][]{kwm88,th85}, and super star clusters within nearby blue compact dwarfs \citep[e.g.,][]{thb98,kj99}.  
And while these studies focus on the free-free emission from galaxies, each was conducted at frequencies $\la$30\,GHz.  
With recent improvements to the backends of existing radio telescopes, such as the Caltech Continuum Backend (CCB) on the Robert C. Byrd Green Bank Telescope (GBT) and the Wideband Interferometric Digital ARchitecture (WIDAR) correlator on the Karl G. Jansky Very Large Array (VLA), the availability of increased bandwidth is making it possible to conduct investigations for large samples of objects at frequencies $\sim$30\,GHz.  

In a recent paper, we presented 33\,GHz photometry taken with the CCB on the GBT as part of the Star Formation in Radio Survey \citep[SFRS;][]{ejm12b}.  
%The GBT single-dish photometry 
%Here, we present 
Building on that work, we obtained 33\,GHz imaging for the SFRS using the VLA, allowing us to map the 33\,GHz emission from each region on $\gtrsim$2" scales, compared to the $\approx$25\arcsec~single-beam GBT photometry.   
These galaxies, which are included in the {\it Spitzer} Infrared Nearby Galaxies Survey \citep[SINGS;][]{rck03} and Key Insights on Nearby Galaxies: a Far-Infrared Survey with {\it Herschel} \citep[KINGFISH;][]{kf11} legacy programs, are well studied and have a wealth of ancillary data available.  
We are currently in the process of reducing and imaging complementary interferometric observations at matched resolution in the S- ($2-4$\,GHz) and Ku- ($12-18$\,GHz) bands (VLA/13B-215; PI. Murphy), which will allow us to extend this analysis by making spectral index maps and doing proper thermal/non-thermal decompositions.  
The complete multi-band survey data and associated full analysis will be presented in a forthcoming paper.  

In this paper, we present catalogs of 33\,GHz images and flux density measurements based on VLA observations of the galaxies included in the SFRS.  
%This paper, which focusses on the highest frequency (33\,GHz) VLA SFRS observations is organized as follows:   
The paper is organized as follows:
In $\S$2 we describe our sample selection and the data used in the present study.  
In $\S$3 we describe our analysis procedures.  
Our results are presented and discussed in $\S$4.  % and discussed in $\S$5.  
Finally, in $\S$5, we summarize our main conclusions.  
Throughout the paper we report median absolute deviations rather than a standard deviations as this statistic is more resilient against outliers in a data set.

\begin{deluxetable}{l|cc}
\tablecaption{Extranuclear Source Positions \label{tbl-2}}
\tabletypesize{\scriptsize}
\tablewidth{0pt}
\tablehead{
\colhead{ID}  & \colhead{R.A.} & \colhead{Decl.} \\
\colhead{} & \colhead{(J2000)} & \colhead{(J2000)}
}
  NGC\,0628~Enuc.\,1   &$  0 1   ~36   ~45.1   $&$+   15   ~47   ~51  $\\
  NGC\,0628~Enuc.\,2   &$  0 1   ~36   ~37.5   $&$+   15   ~45   ~12  $\\
  NGC\,0628~Enuc.\,3   &$  0 1   ~36   ~38.8   $&$+   15   ~44   ~25  $\\
  NGC\,0628~Enuc.\,4   &$  0 1   ~36   ~35.5   $&$+   15   ~50   ~11  $\\
  NGC\,1097~Enuc.\,1   &$  0 2   ~46   ~23.9   $&$-   30   ~17   ~50  $\\
  NGC\,1097~Enuc.\,2   &$  0 2   ~46   ~14.4   $&$-   30   ~15  ~0 4  $\\
  NGC\,2403~Enuc.\,1   &$  0 7   ~36   ~45.5   $&$+   65   ~37  ~0 0  $\\
  NGC\,2403~Enuc.\,2   &$  0 7   ~36   ~52.7   $&$+   65   ~36   ~46  $\\
  NGC\,2403~Enuc.\,3   &$  0 7   ~37  ~0 6.9   $&$+   65   ~36   ~39  $\\
  NGC\,2403~Enuc.\,4   &$  0 7   ~37   ~17.9   $&$+   65   ~33   ~46  $\\
  NGC\,2403~Enuc.\,5   &$  0 7   ~36   ~19.5   $&$+   65   ~37  ~0 4  $\\
  NGC\,2403~Enuc.\,6   &$  0 7   ~36   ~28.5   $&$+   65   ~33   ~50  $\\
  NGC\,2976~Enuc.\,1   &$  0 9   ~47  ~0 7.8   $&$+   67   ~55   ~52  $\\
  NGC\,2976~Enuc.\,2   &$  0 9   ~47   ~24.1   $&$+   67   ~53   ~56  $\\
  NGC\,3521~Enuc.\,1   &$   11  ~0 5   ~46.3   $&$-  0 0  ~0 4  ~0 9  $\\
  NGC\,3521~Enuc.\,2   &$   11  ~0 5   ~49.9   $&$-  0 0  ~0 3   ~39  $\\
  NGC\,3521~Enuc.\,3   &$   11  ~0 5   ~47.6   $&$+  0 0  ~0 0   ~33  $\\
  NGC\,3627~Enuc.\,1   &$   11   ~20   ~16.2   $&$+   12   ~57   ~50  $\\
  NGC\,3627~Enuc.\,2   &$   11   ~20   ~16.3   $&$+   12   ~58   ~44  $\\
  NGC\,3627~Enuc.\,3   &$   11   ~20   ~16.0   $&$+   12   ~59   ~52  $\\
  NGC\,3938~Enuc.\,1   &$   11   ~52   ~46.4   $&$+   44  ~0 7  ~0 1  $\\
  NGC\,3938~Enuc.\,2   &$   11   ~53  ~0 0.0   $&$+   44  ~0 7   ~55  $\\
  NGC\,4254~Enuc.\,1   &$   12   ~18   ~49.1   $&$+   14   ~23   ~59  $\\
  NGC\,4254~Enuc.\,2   &$   12   ~18   ~44.6   $&$+   14   ~24   ~25  $\\
  NGC\,4321~Enuc.\,1   &$   12   ~22   ~58.9   $&$+   15   ~49   ~35  $\\
  NGC\,4321~Enuc.\,2   &$   12   ~22   ~49.8   $&$+   15   ~50   ~29  $\\
  NGC\,4631~Enuc.\,1   &$   12   ~41   ~40.8   $&$+   32   ~31   ~51  $\\
  NGC\,4631~Enuc.\,2   &$   12   ~42   ~21.3   $&$+   32   ~33  ~0 6  $\\
  NGC\,4736~Enuc.\,1   &$   12   ~50   ~56.2   $&$+   41  ~0 7   ~20  $\\
  NGC\,5055~Enuc.\,1   &$   13   ~15   ~58.0   $&$+   42  ~0 0   ~26  $\\
  NGC\,5194~Enuc.\,1   &$   13   ~29   ~53.1   $&$+   47   ~12   ~40  $\\
  NGC\,5194~Enuc.\,2   &$   13   ~29   ~44.1   $&$+   47   ~10   ~21  $\\
  NGC\,5194~Enuc.\,3   &$   13   ~29   ~44.6   $&$+   47  ~0 9   ~55  $\\
  NGC\,5194~Enuc.\,4   &$   13   ~29   ~56.2   $&$+   47   ~14  ~0 7  $\\
  NGC\,5194~Enuc.\,5   &$   13   ~29   ~59.6   $&$+   47   ~14  ~0 1  $\\
  NGC\,5194~Enuc.\,6   &$   13   ~29   ~39.5   $&$+   47  ~0 8   ~35  $\\
  NGC\,5194~Enuc.\,7   &$   13   ~30  ~0 2.5   $&$+   47  ~0 9   ~52  $\\
  NGC\,5194~Enuc.\,8   &$   13   ~30  ~0 1.6   $&$+   47   ~12   ~52  $\\
  NGC\,5194~Enuc.\,9   &$   13   ~29   ~59.9   $&$+   47   ~11   ~12  $\\
 NGC\,5194~Enuc.\,10   &$   13   ~29   ~56.7   $&$+   47   ~10   ~46  $\\
 NGC\,5194~Enuc.\,11   &$   13   ~29   ~49.7   $&$+   47   ~13   ~29  $\\
  NGC\,5457~Enuc.\,1   &$   14  ~0 3   ~10.2   $&$+   54   ~20   ~57  $\\
  NGC\,5457~Enuc.\,2   &$   14  ~0 2   ~55.0   $&$+   54   ~22   ~26  $\\
  NGC\,5457~Enuc.\,3   &$   14  ~0 3   ~41.3   $&$+   54   ~19  ~0 4  $\\
  NGC\,5457~Enuc.\,4   &$   14  ~0 3   ~53.1   $&$+   54   ~22  ~0 6  $\\
  NGC\,5457~Enuc.\,5   &$   14  ~0 3  ~0 1.1   $&$+   54   ~14   ~28  $\\
  NGC\,5457~Enuc.\,6   &$   14  ~0 2   ~28.1   $&$+   54   ~16   ~26  $\\
  NGC\,5457~Enuc.\,7   &$   14  ~0 4   ~29.3   $&$+   54   ~23   ~46  $\\
  NGC\,5713~Enuc.\,1   &$   14   ~40   ~12.1   $&$-  0 0   ~17   ~47  $\\
  NGC\,5713~Enuc.\,2   &$   14   ~40   ~10.5   $&$-  0 0   ~17   ~47  $\\
  NGC\,6946~Enuc.\,1   &$   20   ~35   ~16.6   $&$+   60   ~10   ~57  $\\
  NGC\,6946~Enuc.\,2   &$   20   ~35   ~25.1   $&$+   60   ~10  ~0 3  $\\
  NGC\,6946~Enuc.\,3   &$   20   ~34   ~52.2   $&$+   60   ~12   ~41  $\\
  NGC\,6946~Enuc.\,4   &$   20   ~34   ~19.4   $&$+   60   ~10  ~0 9  $\\
  NGC\,6946~Enuc.\,5   &$   20   ~34   ~39.0   $&$+   60  ~0 4   ~53  $\\
  NGC\,6946~Enuc.\,6   &$   20   ~35  ~0 6.0   $&$+   60   ~11  ~0 0  $\\
  NGC\,6946~Enuc.\,7   &$   20   ~35   ~11.2   $&$+   60  ~0 8   ~59  $\\
  NGC\,6946~Enuc.\,8   &$   20   ~34   ~32.2   $&$+   60   ~10   ~19  $\\
  NGC\,6946~Enuc.\,9   &$   20   ~35   ~12.7   $&$+   60  ~0 8   ~52  $\\
  NGC\,7793~Enuc.\,1   &$   23   ~57   ~48.8   $&$-   32   ~36   ~58  $\\
  NGC\,7793~Enuc.\,2   &$   23   ~57   ~56.1   $&$-   32   ~35   ~40  $\\
  NGC\,7793~Enuc.\,3   &$   23   ~57   ~48.8   $&$-   32   ~34   ~52  $  
\enddata
\end{deluxetable}

\section{Sample and Data Analysis}
In this section we describe the sample selection.  
We additionally present the VLA observations along with our reduction and imaging procedures, and provide a description of the ancillary data utilized for the present study.  

\subsection{Sample Selection}
\label{sec-sample}
The Star Formation in Radio Survey (SFRS) sample comprises nuclear and extranuclear star-forming regions in 56 nearby galaxies ($d < 30$\,Mpc) observed as part of the SINGS \citep{rck03} and KINGFISH \citep{kf11} legacy programs.  
%SINGS\footnote[1]{The {\it Spitzer} Infrared Nearby Galaxies Survey, \citep{rck03}} and KINGFISH\footnote[2]{Key Insights on Nearby Galaxies: a Far-Infrared Survey with {\it Herschel}, \citep{kf11}} legacy programs.  
Each of these nuclear and extranuclear star-forming complexes have mid-infrared [i.e., low resolution from $5-14\,\mu$m ($0\farcm3 \times 0\farcm9$) and high resolution from $10-37\,\mu$m ($0\farcm3 \times 0\farcm4$)] spectral mappings carried out by the IRS instrument on board {\it Spitzer}, and $47\arcsec \times 47\arcsec$ sized {\it Herschel}/PACS far-infrared spectral mappings for a combination of the principal atomic ISM cooling lines of [OI]63\,$\mu$m, [OIII]88\,$\mu$m, [NII]122,205\,$\mu$m, and [CII]158\,$\mu$m.  
NGC\,5194 and NGC\,2403 are exceptions; these galaxies were part of the SINGS sample, but are not formally included in KINGFISH.  
They were observed with {\it Herschel} as part of the Very Nearby Galaxy Survey (VNGS; PI: C. Wilson).  
Similarly, there are additional KINGFISH galaxies that were not part of SINGS, but have existing {\it Spitzer} data:  NGC\,5457 (M\,101), IC\,342, NGC\,3077, and NGC\,2146.  
%SFR ($\lesssim10^{-4}$ to $10~M_{\odot}\,{\rm yr}^{-1}$),

SINGS and KINGFISH galaxies were chosen to cover the full range of integrated properties and ISM conditions found in the local Universe, spanning the full range in morphological types, a factor of $\sim10^{5}$ in infrared (IR: $8-1000\,\mu$m) luminosity, a factor of $\sim10^{3}$ in $L_{\rm IR}/L_{\rm opt}$, and a large range in star formation rate ($\lesssim10^{-3} - 10\,M_{\odot}\,{\rm yr}^{-1}$).  
Similarly, spectroscopically targeted extranuclear sources included in SINGS and KINGFISH were selected to cover the full range of physical conditions and spectral characteristics found in (bright) infrared sources in nearby galaxies, requiring optical and infrared selections.  
Optically selected extranuclear regions were chosen to span a large range in physical properties, including the extinction-corrected production rate of ionizing photons [$Q(H^{0}) \sim10^{49}-10^{52}$ s$^{-1}$], metallicity ($\sim 0.1-3\,Z_{\odot}$), visual extinction ($A_{V} \la 4$\,mag), radiation field intensity (100-fold range), ionizing stellar temperature ($T_{\rm eff} \sim 3.5 - 5.5\times10^{4}$\,K), and local H$_{2}$/H{\sc i} ratios ($\lesssim 0.1 - \gtrsim10$).  
A sub-sample of infrared-selected extranuclear targets were chosen to span a range in $f_{\nu}(8\,\mu{\rm m})/f_{\nu}(24\,\mu{\rm m})$ and $f_{\rm H\alpha}/f_{\nu}(8\,\mu{\rm m})$ ratios.  

The total set of observations over the entire sky consists of 118 star-forming complexes (56 nuclei and 62 extranuclear regions), 112 of which (50 nuclei and 62 extranuclear regions; see Tables \ref{tbl-1} and \ref{tbl-2}, respectively) are observable with the VLA (i.e., having $\delta > -35\degr$).  
%IS THIS CORRECT?  CHECKED WITH DANNY... WHAT WAS SUBMITTED MIGHT HAVE BEEN UPDATED FROM THE FINAL LIST HE GAVE ME, BUT BOTH CALLED FINAL?
The coordinates given in both tables are the VLA pointing centers, which correspond to the centers of the {\it Spitzer} mid-infrared and {\it Herschel} far-infrared spectral line maps.    
Galaxy morphologies, adopted distances, optically-defined nuclear types, diameters ($D_{25}$), inclinations ($i$), and position angles (P.A.) are given in Table \ref{tbl-1}.
When categorizing nuclear types using \citet{hfs97}, we assign them to be star-forming (SF) is they were given an H{\sc ii} classification or AGN if they were given either a Seyfert or LINER classification.  
Galaxy morphologies, diameters, and position angles were taken from the Third Reference Catalog of Bright Galaxies \citep[RC3;][]{rc3}.
For a number of sources, position angles were not given in the RC3 catalog, so we instead use those derived using 2.2\,$\mu$m ($K_{s}$ band) photometry from the Two Micron All Sky Survey (2MASS) and given in \citet{2massLGA}.  
These sources are identified in Table \ref{tbl-1}.    
We calculate inclinations using the method described by \citet{dd97}
such that, 
\begin{equation}
\cos^{2}i = \frac{(b/a)^{2} - (b/a)^{2}_{\rm int}}{1-(b/a)^{2}_{\rm
    int}},
\end{equation}
where $a$ and $b$ are the observed semi-major and semi-minor axes and
the disks are oblate spheroids with an intrinsic axial
ratio \((b/a)_{\rm int} \simeq 0.2\) for morphological types earlier
than Sbc and \((b/a)_{\rm int} \simeq 0.13\) otherwise.  
  
%In Figure \ref{fig-gbtobs}, we show the location of each target region on {\it Spitzer} 24\,$\mu$m images; 
%the corresponding circle diameters are 25\arcsec, which match the FWHM of our lowest resolution data (i.e., the beam of the GBT 33\,GHz radio data) for the present multiwavelength study.  

\subsection{33\,GHz VLA Observations and Data Reduction}
\label{sec-vlaobs}
Observations in the Ka band ($26.5-40$\,GHz) were taken during two separate VLA D-configuration cycles.  
As with our GBT program, the observing strategy was constructed to make the most efficient use of the telescope.  
Thus, given the large range in brightness among our targeted regions, we varied the time spent on source based on an estimate of the expected 33\,GHz flux density using the {\it Spitzer} 24\,$\mu$m maps.  

D-configuration observations were obtained in November 2011 (VLA/11B-032) and March 2013 (VLA/13A-129).
For the first round of observations, the 8-bit samplers were used, yielding 2\,GHz of simultaneous bandwidth, which we used to center 1\,GHz wide basebands at 32.5 and 33.5\,GHz. 
For the latter run, the 3-bit samplers became available, yielding 8\,GHz of instantaneous bandwidth in  2\,GHz wide basebands centered at 30, 32, 34, and 36\,GHz. 
The standard VLA flux density calibrators 3C\,48, 3C\,286, and 3C\,147 were used.  

During the 11B semester, there was a correlator malfunction such that only the first second of all correlator integration times were recorded.  
In our case, we used a 3\,s dump time, resulting in only obtaining $\onethird$ of the requested data.  
Because of this, a fraction of sources included in VLA/11B-032 were re-observed later in the semester, some of which were observed during the move in DnC-configuration.  
We additionally re-observed a number of sources during 13A that were not re-observed in 11B.  
These various cases are identified in Tables \ref{tbl-3} and \ref{tbl-4}
%{\bf Give calibrators?\\
%In both cases, 3C\,286 was used as the flux density and bandpass calibrator, while J\,1118+1234 was used as the complex gain and telescope pointing calibrator. 
%Describe 11B-032 data, including 1\,s issue. \\
%Describe 13A-129 data, and how we use 3-bit correlator setting.  \\
%}

To reduce the VLA data, we used a number of Common Astronomy Software Applications \citep[CASA;][]{casa} versions and followed standard calibration and editing procedures, including the utilization of the VLA calibration pipeline. 
For data calibrated with the VLA pipeline using CASA 4.4.0 or later, we inspected the visibilities and calibration tables for evidence of bad antennas, frequency ranges, and time ranges, flagging correspondingly. 
We also flagged any instances of RFI, for which we found very little of at 33\,GHz. 
After flagging, we re-ran the pipeline, and repeated this process until all bad data was removed.

For data calibrated without the pipeline, we used the following general procedure, and re-generated all previous calibration tables as necessary if antennas, frequencies, or time ranges were flagged for having bad data:
\begin{enumerate}
\item Generate initial calibration tables for antenna position, opacity, and gain curve. 
\item Set the flux calibrator's flux scale using the 2010 version of the  Perley \& Butler model.  %(QUESTION: WAS ALL BY-HAND DATA USING THIS MODEL?)
\item Generate the initial delay calibration table (using the flux calibrator), applying prior calibration tables on-the-fly. 
\item Generate the initial short (15\,s) integration phase-only gain calibration table (using the flux calibrator), applying the delay table and prior tables on-the-fly. 
\item Generate the initial bandpass calibration (using the flux calibrator), applying the delay, phase, and prior tables on-the-fly. 
\item Generate the final short integration phase-only gain calibration tables for all calibrators (flux and phase), applying the delay, bandpass, and prior tables on-the-fly. 
\item Generate amplitude$ + $phase gain calibration tables for all calibrators, applying the delay, short phase, bandpass, and prior tables on-the-fly. 
\item Use the amplitude$ + $phase calibration tables to set the final flux scale calibration for all phase calibrators. 
\item Generate the final long (full scan) integration phase gain calibrations for all calibrators, applying the delay, bandpass, flux scale, and prior tables on-the-fly. 
\item Apply the long integration phase, bandpass, delay, flux scale, and prior tables to all science targets. 
\end{enumerate}
For all delay and bandpass tables applied on-the-fly, we used the default nearest-neighbor interpolation. 
For phase and flux scale tables, we used a linear interpolation.

For all 87 nuclear and extranuclear regions that were calibrated by hand using CASA versions 4.2.1 or earlier, the 2010 Perley \& Butler flux density scale was applied as the default.  
This is different than the flux density scale used in the pipeline calibrated data run with CASA version 4.4.0 or later \citep[i.e., ][]{pb13fcal}.
To place everything on the same flux density scale, we corrected the amplitude of the final images for all 87 regions by multiplying them by the ratio of the Perley \& Butler 2013 to 2010 flux density scalings.  
The average correction factor was near unity at 0.98 with an rms scatter of 0.01.

%For the 44 nuclear and extranuclear regions in the 11B-032 dataset that were reduced without the pipeline, CASA version 3.4.0 was used, which utilized a 2010 version of the Perley \& Butler flux density scale.
%This is different than the flux density scale used in CASA version 4.1.0 or later \citep[i.e., ][]{pb13fcal}.
%To place everything on the same flux density scale, we corrected the amplitude of the final images for these 44 sources by multiplying them by the ratio of the Perley \& Butler 2013 to 2010 flux density scalings.  %(PB2013 Flux Scaling/PB2010 Flux Scaling).
%The average correction factor was near unity at 0.98 with an rms scatter of 0.004.

\subsection{Interferometric Imaging}
Calibrated VLA measurement sets for each source were imaged using the task {\sc tclean} in CASA version 4.6.0.  
For some cases (see Tables \ref{tbl-3} and \ref{tbl-4}), the Ka-band images contain data from observations taken during both the 11B and 13A semesters, but are heavily weighted by the 13A semester observations as those include significantly more data.  
The mode of {\sc tclean} was set to multi-frequency synthesis \citep[{\sc mfs};][]{mfs1,mfs2}.  
We chose to use {\it Briggs} weighting with {\sc robust=0.5}, and set the variable {\sc nterms=2}, which allows the cleaning procedure to also model the spectral index variations on the sky.
To help deconvolve extended low-intensity emission, we took advantage of the multiscale clean option \citep{msclean,msmfs} in CASA, searching for structures with scales $\approx$1 and 3 times the FWHM of the synthesized beam. 
The choice of our final imaging parameters was the result of extensive experimentation to identify values that yielded the best combination of brightness-temperature sensitivity and reduction of artifacts resulting from strong sidelobes in the naturally weighted beam for these snapshot-like observations.  

The images were placed on a $512\times512$ pixel grid with a pixel scale of 0\farcs3.  
However, for two sources (NGC\,0628\,Enuc.\,4 and NGC\,0855), the pixel scale was reduced to 0\farcs15 to ensure that the FWHM of the synthesized beam minor axis remained Nyquist sampled.  

For two sources in the sample, NGC\,4594 and NGC\,4579, a signal-to-noise ratio ($SNR$) $\geq 3$ was achieved across the majority of all channels and spectral windows. 
This allowed us to accurately perform phase only, and subsequently amplitude$ + $phase, self-calibration for these two sources.
The peak brightness of the self-calibrated images differs from that of the originals by less than 5\%, however the new peak $SNRs$ of NGC\,4594 and NGC\,4579 are improved by factors of $\approx$2 and $\approx$3, respectively (achieving peak $SNRs$ of $\sim$2900 and $\sim$1500, respectively).

A primary beam correction was applied using the CASA task {\sc impbcor} before analyzing the images.
The primary-beam-corrected continuum images at 33\,GHz for each target are shown in Figure \ref{fig:imgs}.   
The FWHM of the synthesized beams are given in Tables \ref{tbl-3} and \ref{tbl-4} for all sources, along with the corresponding point-source and brightness temperature rms values for each of the final images.  
%he FWHM of the synthesized beams ranged between $\theta_{\rm maj} \approx 1.7-4.3\arcsec$, depending on the source declination (see Tables \ref{tbl-1} and \ref{tbl-2}).  
Given the range of distances to the sample galaxies, this ensured that the linear scale investigated was always $\la$300\,pc (i.e., the size of giant H{\sc ii} regions).  
%We also note that the largest angular scale that the VLA images should be sensitive to is $\approx$44\arcsec.  
We also note that the VLA images made with the chosen array configurations should be sensitive to extended emission on angular scales up to $\approx$24\arcsec~for these snapshot observations.  

We also created a suite of $(u, v)$-tapered images for all nuclear and extranuclear regions in order to assess the potential for missing large-scale emission. 
After tapering to 2\farcs5, we find that we recover $\sim 3\%$ more flux density relative to the non-tapered images, suggesting that on the scales of the individual H{\sc ii} regions and nuclei, we are not missing a significant amount of the source flux density.  
%are native resoluiton images are not missi when comparing to the total flux density at 33 GHz as measured with the GBT.

\begin{deluxetable*}{l|cccc}
\tablecaption{Nuclear Source Imaging Characteristics \label{tbl-3}}
\tabletypesize{\scriptsize}
\tablewidth{0pt}
\tablehead{
\colhead{Galaxy}  & \colhead{Program ID} & \colhead{Synthesized}& \colhead{$\sigma$}& \colhead{$\sigma_{T_{\rm b}}$}\\
\colhead{}  & \colhead{} & \colhead{Beam}& \colhead{($\mu$Jy\,bm$^{-1}$)}& \colhead{(mK)}
}
           NGC\,0337   &                 VLA/11B-32\tablenotemark{b}  &  $2\farcs04 \times 1\farcs13$  &12.5  & 6.05\\
           NGC\,0628   &                 VLA/11B-32\tablenotemark{a}  &  $2\farcs14 \times 1\farcs94$  &21.1  & 5.66\\
           NGC\,0855   &                 VLA/11B-32\tablenotemark{b}  &  $1\farcs92 \times 0\farcs93$  &11.3  & 7.12\\
           NGC\,0925   &                 VLA/11B-32\tablenotemark{b}  &  $1\farcs91 \times 1\farcs36$  &12.9  & 5.56\\
           NGC\,1097   &                 VLA/11B-32\tablenotemark{b}  &  $3\farcs17 \times 1\farcs55$  &43.1  & 9.80\\
           NGC\,1266   &                 VLA/11B-32\tablenotemark{a}  &  $2\farcs42 \times 1\farcs88$  &52.2  &12.82\\
           NGC\,1377   &                 VLA/11B-32\tablenotemark{a}  &  $3\farcs57 \times 1\farcs91$  &29.8  & 4.87\\
            IC\,0342   &                 VLA/11B-32\tablenotemark{b}  &  $1\farcs75 \times 1\farcs72$  &34.6  &12.78\\
           NGC\,1482   &                 VLA/11B-32\tablenotemark{a}  &  $3\farcs27 \times 1\farcs79$  &72.0  &13.78\\
           NGC\,2146   &                 VLA/11B-32\tablenotemark{b}  &  $1\farcs90 \times 1\farcs07$  &34.1  &18.79\\
           NGC\,2403   &                                 VLA/13A-129  &  $2\farcs42 \times 1\farcs85$  & 9.9  & 2.47\\
        Holmberg\,II   &                 VLA/11B-32\tablenotemark{b}  &  $1\farcs84 \times 1\farcs01$  &14.9  & 8.92\\
           NGC\,2798   &                 VLA/11B-32\tablenotemark{a}  &  $2\farcs07 \times 1\farcs74$  &18.1  & 5.63\\
           NGC\,2841   &                 VLA/11B-32\tablenotemark{a}  &  $2\farcs08 \times 1\farcs89$  &10.1  & 2.85\\
           NGC\,2976   &                 VLA/11B-32\tablenotemark{b}  &  $2\farcs40 \times 1\farcs68$  &19.7  & 5.46\\
           NGC\,3049   &    VLA/11B-32,\tablenotemark{a}~VLA/13A-129  &  $2\farcs50 \times 2\farcs02$  &17.9  & 3.95\\
           NGC\,3077   &                 VLA/11B-32\tablenotemark{b}  &  $2\farcs45 \times 1\farcs66$  &29.3  & 8.05\\
           NGC\,3190   &    VLA/11B-32,\tablenotemark{a}~VLA/13A-129  &  $2\farcs13 \times 1\farcs85$  &13.7  & 3.86\\
           NGC\,3184   &    VLA/11B-32,\tablenotemark{a}~VLA/13A-129  &  $2\farcs51 \times 1\farcs93$  &13.1  & 3.02\\
           NGC\,3198   &                 VLA/11B-32\tablenotemark{a}  &  $2\farcs06 \times 1\farcs98$  &18.7  & 5.11\\
            IC\,2574   &                 VLA/11B-32\tablenotemark{b}  &  $2\farcs17 \times 1\farcs64$  &15.7  & 4.92\\
           NGC\,3265   &    VLA/11B-32,\tablenotemark{a}~VLA/13A-129  &  $2\farcs15 \times 1\farcs94$  &12.9  & 3.44\\
           NGC\,3351   &    VLA/11B-32,\tablenotemark{a}~VLA/13A-129  &  $2\farcs27 \times 2\farcs04$  &17.6  & 4.24\\
           NGC\,3521   &                 VLA/11B-32\tablenotemark{a}  &  $4\farcs20 \times 1\farcs98$  &27.3  & 3.66\\
           NGC\,3621   &                 VLA/11B-32\tablenotemark{a}  &  $4\farcs32 \times 1\farcs58$  &30.3  & 4.96\\
           NGC\,3627   &    VLA/11B-32,\tablenotemark{a}~VLA/13A-129  &  $2\farcs83 \times 1\farcs83$  &23.7  & 5.10\\
           NGC\,3773   &    VLA/11B-32,\tablenotemark{a}~VLA/13A-129  &  $2\farcs99 \times 2\farcs50$  &20.3  & 3.03\\
           NGC\,3938   &                 VLA/11B-32\tablenotemark{b}  &  $2\farcs25 \times 1\farcs83$  &16.5  & 4.47\\
           NGC\,4254   &                                 VLA/13A-129  &  $2\farcs34 \times 1\farcs90$  &15.4  & 3.87\\
           NGC\,4321   &                                 VLA/13A-129  &  $2\farcs41 \times 1\farcs77$  &18.4  & 4.83\\
           NGC\,4536   &                                 VLA/13A-129  &  $2\farcs36 \times 2\farcs16$  &17.4  & 3.79\\
           NGC\,4559   &                                 VLA/13A-129  &  $3\farcs02 \times 1\farcs94$  &11.1  & 2.11\\
           NGC\,4569   &                                 VLA/13A-129  &  $2\farcs41 \times 1\farcs76$  &21.7  & 5.70\\
           NGC\,4579   &                                 VLA/13A-129  &  $2\farcs48 \times 1\farcs78$  &34.2  & 8.65\\
           NGC\,4594   &                                 VLA/13A-129  &  $2\farcs98 \times 2\farcs12$  &20.2  & 3.57\\
           NGC\,4625   &                                 VLA/13A-129  &  $2\farcs96 \times 2\farcs09$  & 9.2  & 1.66\\
           NGC\,4631   &                                 VLA/13A-129  &  $2\farcs33 \times 1\farcs97$  &15.7  & 3.82\\
           NGC\,4725   &                                 VLA/13A-129  &  $2\farcs89 \times 1\farcs97$  &10.7  & 2.09\\
           NGC\,4736   &                                 VLA/13A-129  &  $2\farcs98 \times 2\farcs09$  &20.5  & 3.68\\
           NGC\,4826   &                                 VLA/13A-129  &  $2\farcs16 \times 1\farcs98$  &14.0  & 3.64\\
           NGC\,5055   &                                 VLA/13A-129  &  $2\farcs76 \times 2\farcs12$  &16.9  & 3.23\\
           NGC\,5194   &                                 VLA/13A-129  &  $2\farcs27 \times 1\farcs80$  &13.6  & 3.72\\
           NGC\,5398   &                                 VLA/13A-129  &  $5\farcs42 \times 1\farcs79$  &18.1  & 2.08\\
           NGC\,5457   &                                 VLA/13A-129  &  $2\farcs36 \times 1\farcs76$  &14.1  & 3.79\\
           NGC\,5474   &                                 VLA/13A-129  &  $2\farcs18 \times 1\farcs84$  & 9.4  & 2.61\\
           NGC\,5713   &                                 VLA/13A-129  &  $2\farcs39 \times 2\farcs14$  &14.5  & 3.17\\
           NGC\,5866   &                                 VLA/13A-129  &  $2\farcs28 \times 1\farcs79$  &16.2  & 4.43\\
           NGC\,6946   &                 VLA/11B-32\tablenotemark{b}  &  $2\farcs12 \times 1\farcs70$  &31.5  & 9.72\\
           NGC\,7331   &                 VLA/11B-32\tablenotemark{a}  &  $3\farcs02 \times 1\farcs88$  &35.8  & 7.04\\
           NGC\,7793   &                 VLA/11B-32\tablenotemark{a}  &  $4\farcs48 \times 1\farcs69$  &23.6  & 3.48  
\enddata
\tablecomments{VLA11B-32 observations were conducted between October 2011 and January 2012.  VLA/13A-129 observations were conducted between Februrary and March 2013.}
\tablenotetext{a}{Observations suffered from the ``1\,s" WIDAR correlator malfunction, leading to only \onethird~of the integration time being recorded.}
\tablenotetext{b}{Original observations suffered from the ``1\,s" WIDAR correlator malfunction, but the source was later reobserved for the nominal integration time.}
\end{deluxetable*}

\begin{deluxetable*}{l|cccc}
\tablecaption{Extranuclear Source Imaging Characteristics \label{tbl-4}}
\tabletypesize{\scriptsize}
\tablewidth{0pt}
\tablehead{
\colhead{Galaxy}  & \colhead{Program ID} & \colhead{Synthesized}& \colhead{$\sigma$}& \colhead{$\sigma_{T_{\rm b}}$}\\
\colhead{}  & \colhead{} & \colhead{Beam}& \colhead{($\mu$Jy\,bm$^{-1}$)}& \colhead{(mK)}
}
  NGC\,0628~Enuc.\,1   &                 VLA/11B-32\tablenotemark{a}  &  $2\farcs08 \times 1\farcs92$  &25.7  & 7.20\\
  NGC\,0628~Enuc.\,2   &                 VLA/11B-32\tablenotemark{a}  &  $2\farcs04 \times 1\farcs88$  &19.7  & 5.73\\
  NGC\,0628~Enuc.\,3   &                 VLA/11B-32\tablenotemark{a}  &  $2\farcs05 \times 1\farcs80$  &26.6  & 8.04\\
  NGC\,0628~Enuc.\,4   &                 VLA/11B-32\tablenotemark{b}  &  $1\farcs75 \times 0\farcs94$  &10.8  & 7.32\\
  NGC\,1097~Enuc.\,1   &                 VLA/11B-32\tablenotemark{b}  &  $2\farcs00 \times 1\farcs71$  &13.7  & 4.47\\
  NGC\,1097~Enuc.\,2   &                 VLA/11B-32\tablenotemark{b}  &  $2\farcs11 \times 1\farcs67$  &14.1  & 4.48\\
  NGC\,2403~Enuc.\,1   &                                 VLA/13A-129  &  $2\farcs60 \times 1\farcs80$  &14.1  & 3.36\\
  NGC\,2403~Enuc.\,2   &                                 VLA/13A-129  &  $2\farcs54 \times 1\farcs80$  &13.8  & 3.38\\
  NGC\,2403~Enuc.\,3   &                                 VLA/13A-129  &  $2\farcs52 \times 1\farcs79$  &18.1  & 4.47\\
  NGC\,2403~Enuc.\,4   &                                 VLA/13A-129  &  $2\farcs44 \times 1\farcs81$  &10.0  & 2.53\\
  NGC\,2403~Enuc.\,5   &                                 VLA/13A-129  &  $2\farcs75 \times 1\farcs79$  &13.9  & 3.15\\
  NGC\,2403~Enuc.\,6   &                                 VLA/13A-129  &  $2\farcs71 \times 1\farcs83$  & 9.8  & 2.21\\
  NGC\,2976~Enuc.\,1   &                 VLA/11B-32\tablenotemark{b}  &  $2\farcs39 \times 1\farcs66$  &19.6  & 5.53\\
  NGC\,2976~Enuc.\,2   &                 VLA/11B-32\tablenotemark{b}  &  $2\farcs38 \times 1\farcs71$  &21.7  & 5.93\\
  NGC\,3521~Enuc.\,1   &                 VLA/11B-32\tablenotemark{a}  &  $4\farcs01 \times 2\farcs08$  &36.8  & 4.94\\
  NGC\,3521~Enuc.\,2   &                 VLA/11B-32\tablenotemark{a}  &  $4\farcs74 \times 1\farcs93$  &34.3  & 4.18\\
  NGC\,3521~Enuc.\,3   &                 VLA/11B-32\tablenotemark{a}  &  $4\farcs23 \times 1\farcs95$  &26.8  & 3.63\\
  NGC\,3627~Enuc.\,1   &    VLA/11B-32,\tablenotemark{a}~VLA/13A-129  &  $2\farcs45 \times 2\farcs03$  &19.6  & 4.40\\
  NGC\,3627~Enuc.\,2   &    VLA/11B-32,\tablenotemark{a}~VLA/13A-129  &  $2\farcs55 \times 2\farcs08$  &19.4  & 4.08\\
  NGC\,3627~Enuc.\,3   &    VLA/11B-32,\tablenotemark{a}~VLA/13A-129  &  $2\farcs43 \times 1\farcs93$  &14.3  & 3.41\\
  NGC\,3938~Enuc.\,1   &                 VLA/11B-32\tablenotemark{b}  &  $2\farcs34 \times 1\farcs83$  &19.4  & 5.04\\
  NGC\,3938~Enuc.\,2   &                 VLA/11B-32\tablenotemark{b}  &  $2\farcs22 \times 1\farcs78$  &21.0  & 5.93\\
  NGC\,4254~Enuc.\,1   &                                 VLA/13A-129  &  $2\farcs35 \times 1\farcs93$  &16.0  & 3.97\\
  NGC\,4254~Enuc.\,2   &                                 VLA/13A-129  &  $2\farcs40 \times 1\farcs96$  &10.8  & 2.56\\
  NGC\,4321~Enuc.\,1   &                                 VLA/13A-129  &  $2\farcs34 \times 1\farcs79$  &12.1  & 3.20\\
  NGC\,4321~Enuc.\,2   &                                 VLA/13A-129  &  $2\farcs33 \times 1\farcs82$  &12.3  & 3.24\\
  NGC\,4631~Enuc.\,1   &                                 VLA/13A-129  &  $2\farcs42 \times 1\farcs99$  &10.6  & 2.46\\
  NGC\,4631~Enuc.\,2   &                                 VLA/13A-129  &  $2\farcs23 \times 1\farcs98$  &11.1  & 2.79\\
  NGC\,4736~Enuc.\,1   &                                 VLA/13A-129  &  $2\farcs89 \times 2\farcs03$  &17.9  & 3.42\\
  NGC\,5055~Enuc.\,1   &                                 VLA/13A-129  &  $2\farcs81 \times 2\farcs05$  &15.3  & 2.97\\
  NGC\,5194~Enuc.\,1   &                                 VLA/13A-129  &  $2\farcs12 \times 1\farcs81$  &14.1  & 4.11\\
  NGC\,5194~Enuc.\,2   &                                 VLA/13A-129  &  $2\farcs32 \times 1\farcs80$  &13.2  & 3.54\\
  NGC\,5194~Enuc.\,3   &                                 VLA/13A-129  &  $2\farcs17 \times 1\farcs80$  & 9.8  & 2.79\\
  NGC\,5194~Enuc.\,4   &                                 VLA/13A-129  &  $2\farcs10 \times 1\farcs84$  &10.2  & 2.96\\
  NGC\,5194~Enuc.\,5   &                                 VLA/13A-129  &  $2\farcs37 \times 1\farcs73$  &19.8  & 5.39\\
  NGC\,5194~Enuc.\,6   &                                 VLA/13A-129  &  $2\farcs39 \times 1\farcs82$  & 9.4  & 2.42\\
  NGC\,5194~Enuc.\,7   &                                 VLA/13A-129  &  $2\farcs38 \times 1\farcs80$  &19.6  & 5.11\\
  NGC\,5194~Enuc.\,8   &                                 VLA/13A-129  &  $2\farcs38 \times 1\farcs80$  &18.5  & 4.83\\
  NGC\,5194~Enuc.\,9   &                                 VLA/13A-129  &  $2\farcs38 \times 1\farcs77$  &20.5  & 5.46\\
 NGC\,5194~Enuc.\,10   &                                 VLA/13A-129  &  $2\farcs07 \times 1\farcs84$  &15.4  & 4.52\\
 NGC\,5194~Enuc.\,11   &                                 VLA/13A-129  &  $2\farcs14 \times 1\farcs81$  & 9.9  & 2.84\\
  NGC\,5457~Enuc.\,1   &                                 VLA/13A-129  &  $2\farcs26 \times 1\farcs80$  & 8.9  & 2.44\\
  NGC\,5457~Enuc.\,2   &                                 VLA/13A-129  &  $2\farcs42 \times 2\farcs33$  &15.2  & 3.02\\
  NGC\,5457~Enuc.\,3   &                                 VLA/13A-129  &  $2\farcs37 \times 2\farcs26$  &21.9  & 4.57\\
  NGC\,5457~Enuc.\,4   &                                 VLA/13A-129  &  $2\farcs28 \times 1\farcs78$  &13.8  & 3.78\\
  NGC\,5457~Enuc.\,5   &                                 VLA/13A-129  &  $2\farcs43 \times 2\farcs33$  &14.8  & 2.92\\
  NGC\,5457~Enuc.\,6   &                                 VLA/13A-129  &  $2\farcs43 \times 2\farcs34$  &15.5  & 3.05\\
  NGC\,5457~Enuc.\,7   &                                 VLA/13A-129  &  $2\farcs21 \times 1\farcs80$  &13.5  & 3.78\\
  NGC\,5713~Enuc.\,1   &                                 VLA/13A-129  &  $2\farcs36 \times 2\farcs16$  &10.5  & 2.31\\
  NGC\,5713~Enuc.\,2   &                                 VLA/13A-129  &  $2\farcs44 \times 2\farcs15$  &14.3  & 3.03\\
  NGC\,6946~Enuc.\,1   &                 VLA/11B-32\tablenotemark{b}  &  $2\farcs08 \times 1\farcs78$  &16.2  & 4.87\\
  NGC\,6946~Enuc.\,2   &                 VLA/11B-32\tablenotemark{b}  &  $2\farcs13 \times 1\farcs87$  &17.3  & 4.85\\
  NGC\,6946~Enuc.\,3   &                 VLA/11B-32\tablenotemark{b}  &  $2\farcs11 \times 1\farcs86$  &10.5  & 3.01\\
  NGC\,6946~Enuc.\,4   &                 VLA/11B-32\tablenotemark{b}  &  $2\farcs17 \times 1\farcs82$  &10.7  & 3.01\\
  NGC\,6946~Enuc.\,5   &                 VLA/11B-32\tablenotemark{b}  &  $2\farcs08 \times 1\farcs86$  &10.8  & 3.12\\
  NGC\,6946~Enuc.\,6   &                 VLA/11B-32\tablenotemark{b}  &  $2\farcs08 \times 1\farcs74$  &16.3  & 5.05\\
  NGC\,6946~Enuc.\,7   &                 VLA/11B-32\tablenotemark{b}  &  $2\farcs08 \times 1\farcs78$  &16.1  & 4.86\\
  NGC\,6946~Enuc.\,8   &                 VLA/11B-32\tablenotemark{b}  &  $2\farcs14 \times 1\farcs70$  &16.7  & 5.09\\
  NGC\,6946~Enuc.\,9   &                 VLA/11B-32\tablenotemark{b}  &  $2\farcs08 \times 1\farcs78$  &16.1  & 4.86\\
  NGC\,7793~Enuc.\,1   &                 VLA/11B-32\tablenotemark{a}  &  $4\farcs25 \times 1\farcs64$  &29.8  & 4.78\\
  NGC\,7793~Enuc.\,2   &                 VLA/11B-32\tablenotemark{a}  &  $4\farcs46 \times 1\farcs70$  &22.5  & 3.32\\
  NGC\,7793~Enuc.\,3   &                 VLA/11B-32\tablenotemark{a}  &  $4\farcs60 \times 1\farcs58$  &37.0  & 5.67  
\enddata
\tablecomments{VLA11B-32 observations were conducted between October 2011 and January 2012.  VLA/13A-129 observations were conducted between Februrary and March 2013.}
\tablenotetext{a}{Observations suffered from the ``1\,s" WIDAR correlator malfunction, leading to only \onethird~of the integration time being recorded.}
\tablenotetext{b}{Original observations suffered from the ``1\,s" WIDAR correlator malfunction, but the source was later reobserved for the nominal integration time.}
\end{deluxetable*}

%\begin{comment}
\subsection{Ancillary Data} %UV, Optical, Infrared, and Radio Data}
\label{sec:ancdata}
%{\bf NEED TO EDIT DOWN TO THE APPROPRIATE DATASETS}\\
The H$\alpha$ imaging used in the analysis is taken from references cited in the compilation by \citet{akl12}, where details about the data quality and preparation (e.g., correction for [NII] emission) can be found.  
H$\alpha$ images were corrected for foreground stars.   
The typical resolution of the seeing-limited H$\alpha$ images is $\approx$1-2\arcsec, and the calibration uncertainty among these maps is taken to be $\approx$20\%.  

Archival {\it Spitzer} 24\,$\mu$m data shown in Figure \ref{fig:imgs} were largely taken from the SINGS and Local Volume Legacy (LVL) legacy programs, and have a calibration uncertainty of $\approx$5\%.    
Details on the associated observation strategies and data reduction steps can be found in \citet{dd07} and \citet{dd09}, respectively.  
Two galaxies, IC\,342 and NGC\,2146, were not a part of SINGS or LVL; their 24\,$\mu$m imaging comes from \citet{ce08}.

\subsection{H$\alpha$ and 33\,GHz Aperture Photometry}

Before making photometric measurements, we aligned the H$\alpha$ images to the 33\,GHz VLA images, which have sub-arcsecond astrometric accuracy. 
In most cases, the H$\alpha$ images had existing astrometric solutions matching multiple H$\alpha$ peaks with 33\,GHz counterparts to better than half of the synthesized beam FWHM.  
%that matched H$\alpha$ peaks with 33 GHz counterparts to less than a $\sim$1\arcsec~offset. 
We adopted the existing astrometry for these galaxies. 
For those remaining galaxies with multiple bright radio sources (e.g., NGC\,0628), we aligned the H$\alpha$ images by eye, ensuring that the peaks of multiple bright features matched their radio counterparts within 1\arcsec.  
While there may be physical offsets between 33\,GHz and H$\alpha$ emission arising from high levels of extinction, we note that these offsets are unlikely to be systematic for multiple distinct peaks. 
In our alignment process, we did not encounter any cases for which the astrometry is significantly affected\footnote{The nucleus of NGC\,4631 is a good example of a case where the H$\alpha$ and 33\,GHz morphologies are clearly distinct. 
For this galaxy, we note that outside of the 33\,GHz field-of-view, the H$\alpha$ and 24\,$\mu$m images align to better than 1\arcsec~and that within the 33\,GHz field-of-view, multiple 24\,$\mu$m and 33\,GHz peaks align to $\approx$0.5\arcsec. 
We suspect that the H$\alpha$ vs 33\,GHz mismatches are caused by high extinction along the line-of-sight into this edge-on galaxy.}.
%While there may be physical offsets between 33\,GHz and H$\alpha$ emission arising from high levels of extinction, we note that these offsets are unlikely to be in the same direction for multiple distinct peaks. 
%Aside from the nucleus of the edge-on galaxy NGC 4631, we did not encounter any cases where the astrometry might be significantly affected.}
%While it is possible that offsets between 33\,GHz and H$\alpha$ emission may be physical, arising from high levels of extinction, we did not encounter any such instances that would affect our astrometry. }
We adopt the existing astrometry
%No alignment was done 
for galaxies with only one detected radio source (e.g., NGC\,3198).

Due to the higher intrinsic brightness of the H$\alpha$ transition relative to free-free emission, our source detection is primarily limited by the 33\,GHz noise and brightness temperature sensitivity given in Tables \ref{tbl-3} and \ref{tbl-4}. 
Because of this, we identified photometric regions by drawing rectangular and polygon apertures around strongly detected 33\,GHz sources. 
Using {\sc PyBDSM}\footnote{\url{http://www.astron.nl/citt/pybdsm/}} \citep{pybdsm}, we have verified that the native resolution 33\,GHz selected sample is complete down to 5$\sigma$ for sources with angular sizes comparable to the $\sim$2\arcsec~synthesized beam.
To minimize the relative contribution from large angular scale H$\alpha$ emission that might fall under the 33\,GHz brightness temperature sensitivity threshold, these apertures are drawn tightly around the brightest parts of the 33\,GHz sources. 
For 33\,GHz non-detections, we simply drew a large aperture encompassing H$\alpha$ (or 24\,$\mu$m, see \S\ref{sec:24umphot}) structures near the phase center. 
%%%[perhaps the rest of this paragraph should go in the caption for the table?] 
The regions, listed in Table~\ref{tbl-5}, are named according to the nearest 33\,GHz image, with an alphabetical suffix if there are multiple regions corresponding to one image. 
For example, ``NGC\,2403\,Enuc\,2.\,B" is one of 3 regions in the image of extranuclear region 2 in NGC\,2403. 
It is also visible in the image of NGC\,2403's nucleus, which has only one (non-detection) region: ``NGC\,2403" (see Figure~\ref{fig:imgs}).

Using the CASA task {\sc imstat}, we measured and report the H$\alpha$ line flux and 33\,GHz flux density for each region in Table \ref{tbl-5} detected with a $SNR >3$.  
For sources that are not detected at this significance we provide a corresponding 3$\sigma$ upper limit.  
The uncertainty in the 33\,GHz flux density is taken to be the standard VLA calibration uncertainty \citep[$\sim$3\%;][]{pb13fcal} added in quadrature with the empirically measured noise from empty regions in each image given in Tables \ref{tbl-3} and \ref{tbl-4}.  % (add in blurb about your iterative method?) 
As stated in \S\ref{sec:ancdata}, the calibration uncertainty of the H$\alpha$ narrowband imaging is $\approx$20\%, which dominates the uncertainty of the H$\alpha$ photometry.  
Also provided in Table \ref{tbl-5} is a measure of the galactocentric radius ($r_\mathrm{G}$) in units of kpc for each position.  
These values are calculated using the assumed galaxy inclinations, position angles, and distances listed in Table \ref{tbl-1}.  

\begin{comment}
\begin{itemize}
\item flux scale errors for H alpha are already discussed in 2.4 Ancillary Data - perhaps move some stuff down here? 

\item Add in a note that there is significant uncertainty in the foreground dust correction for NGC 6946, which is likely a large cause of the negative extinction we find. 

\item Agreed that we shouldn?t give the SFR estimates until we have thermal fractions. Additionally, we should only give extinction estimates for enucs, since most of the nuclei are synchrotron sources.
\end{itemize}
\end{comment}

\subsection{Inclusion of 24\,$\mu$m Data with Aperture Photometry}
\label{sec:24umphot}
To accurately match the photometry obtained with the 33\,GHz and H$\alpha$ images to that measured using the {\it Spitzer} 24\,$\mu$m data, which is at much lower resolution ($\approx 7\arcsec$), we first resolution matched the images.  
Both the 33\,GHz and H$\alpha$ were convolved with a Gaussian kernel resulting in a final FWHM of 7\arcsec.  
Following the image registration method of \citet{ga11}, we convolved the 24\,$\mu$m maps with a kernel that both subtracts out the complex 24\,$\mu$m PSF and restores the image with a Gaussian PSF having a FWHM of 7\arcsec.  

Using the resolution matched images, we measured the flux density within apertures having a 7\arcsec~diameter.   
No attempt was made to apply an aperture correction to the convolved-map photometry as we are only interested in using these data to compare relative values measured at these three bands.  
The majority of these 179 apertures were created by centering a 7\arcsec~diameter circle around the peak pixel in each native resolution aperture and removing apertures that overlap significantly with others or are strongly contaminated by emission from nearby bright sources.  
Additionally, we have created 7\arcsec~apertures for 17 ``diffuse detections" where the 33GHz emission is intrinsically faint and diffuse such that it falls below our compact source detection threshold on 2\arcsec~scales, but constitutes a  $>5\sigma$ detection on 7\arcsec~scales. 

The corresponding 33\,GHz flux densities, H$\alpha$ line fluxes, and 24\,$\mu$m flux densities for each region detected with a $SNR >3$ are given in Table \ref{tbl-6} .
Similar to the naming convention for the photometry carried out at the full resolution of the 33\,GHz maps, sources are named according to the nearest 33\,GHz image, with an alphabetical suffix if there are multiple regions corresponding to one image. 
However, we distinguish individual sources identified in the smoothed maps by instead using a lowercase letter.  
For example, ``NGC\,2403\,Enuc\,2.\,b" is one of two regions in the image of extranuclear region 2 in NGC\,2403, and is composed of the sum contribution of NGC\,2403\,Enuc\,2.\,B and NGC\,2403\,Enuc\,2.\,C in the full-resolution maps.  
For sources that are not detected at this significance we provide a corresponding 3$\sigma$ upper limit.  
As done for sources listed in Table \ref{tbl-5}, we similarly provide a measure of the galactocentric radius in units of kpc for each position.  

\section{Results}

\begin{figure*}[htb!]
\epsscale{1.0}
\plottwo{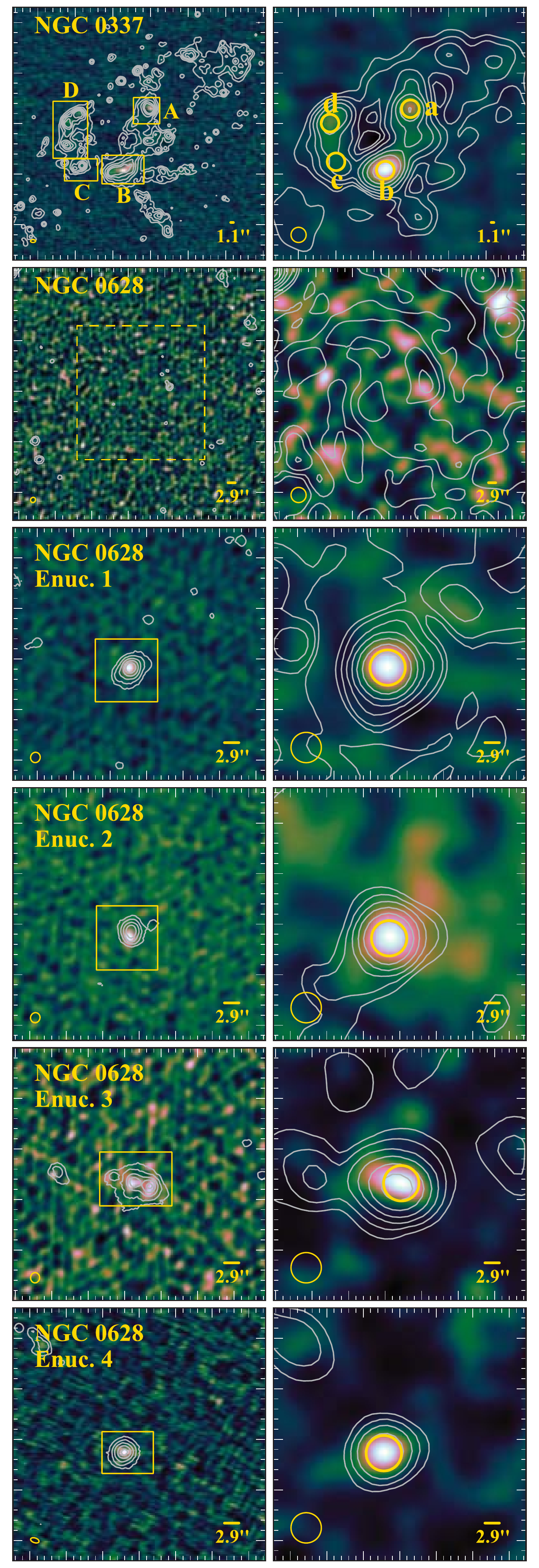}{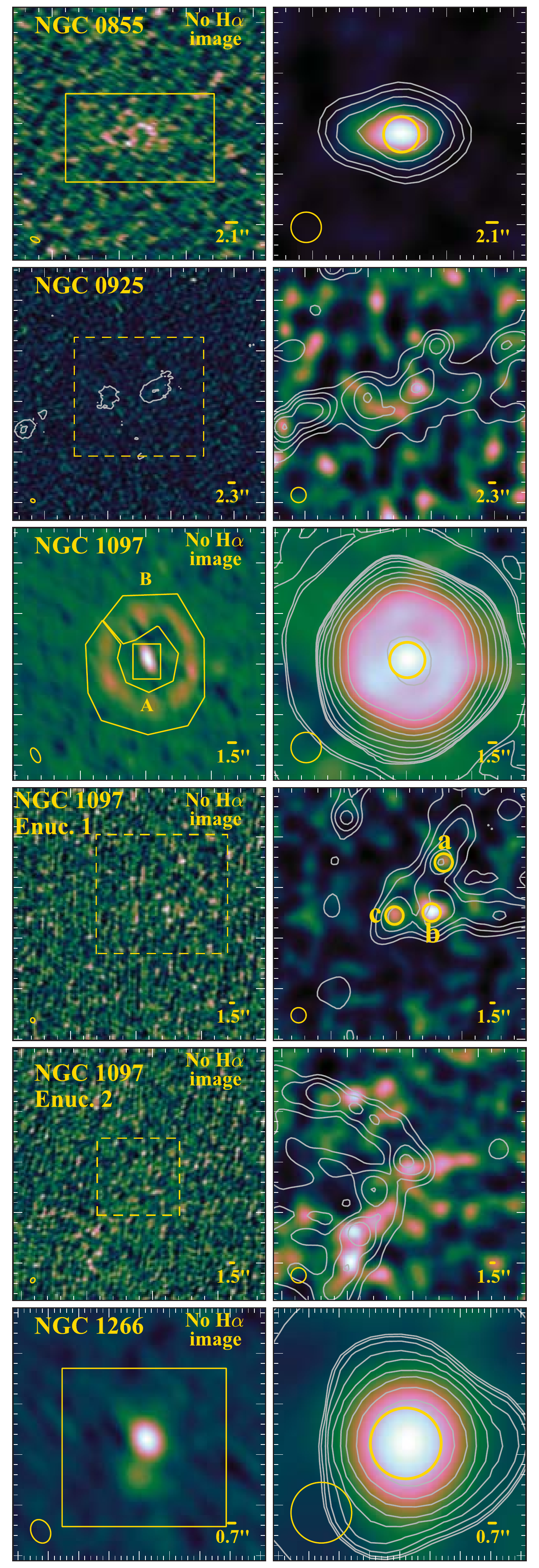}
\caption{ 
%Image cutouts of each target are shown.  
%The 33\,GHz color scale is set to one of 3 power law stretches% (so that the source would stand out more against the noise).
Image cutouts of each target are shown. 
The color scale \citep{green11} is set to one of 3 power law stretches: $[(p-p_{\rm min})/(p_{\rm max} - p_{\rm min})]^a$, where $p$ is the pixel value and $a = 0.5$, 1.0, and 2.0. 
A square-root stretch of $a = 0.5$ was used when the brightest pixel in the image had a $SNR > 20$.  % (to bring out some of the low level emission). 
A linear linear stretch was used when the brightest pixel lied between $10 <  SNR < 20$, and the square stretch was used with the brightest pixel had a $SNR < 10$.  
{\it Left:} The 33\,GHz image at its native (i.e., $\approx$2\arcsec) resolution overlaid with H$\alpha$ contours.  
The H$\alpha$ contours are set at the following values: $[-5,20,40,80,160,320]\sigma$, where $\sigma$ is the local rms noise. 
{\it Right:} The 33\,GHz image convolved to match the resolution of the 24\,$\mu$m data, for which contours are overlaid.   
Depending on the angular size of each source, the cutout regions are either $50\arcsec \times50\arcsec$, $25\arcsec \times 25\arcsec$, or $12\,\farcs5 \times 12\,\farcs5$.  
In all panels, the FWHM of the 33\,GHz beam is shown in the bottom left corner.  
A linear scale-bar of 100\,pc is also given in the bottom right corner of each panel.    
To distinguish between individual sources identified in the full-resolution and smoothed maps, we use upper- and lower-case letters as part of their names for reporting photometry in Table \ref{tbl-5} and \ref{tbl-6}, respectfully. 
\label{fig:imgs}}
\end{figure*}

\begin{figure*}[t!]
\epsscale{1.1}
\plottwo{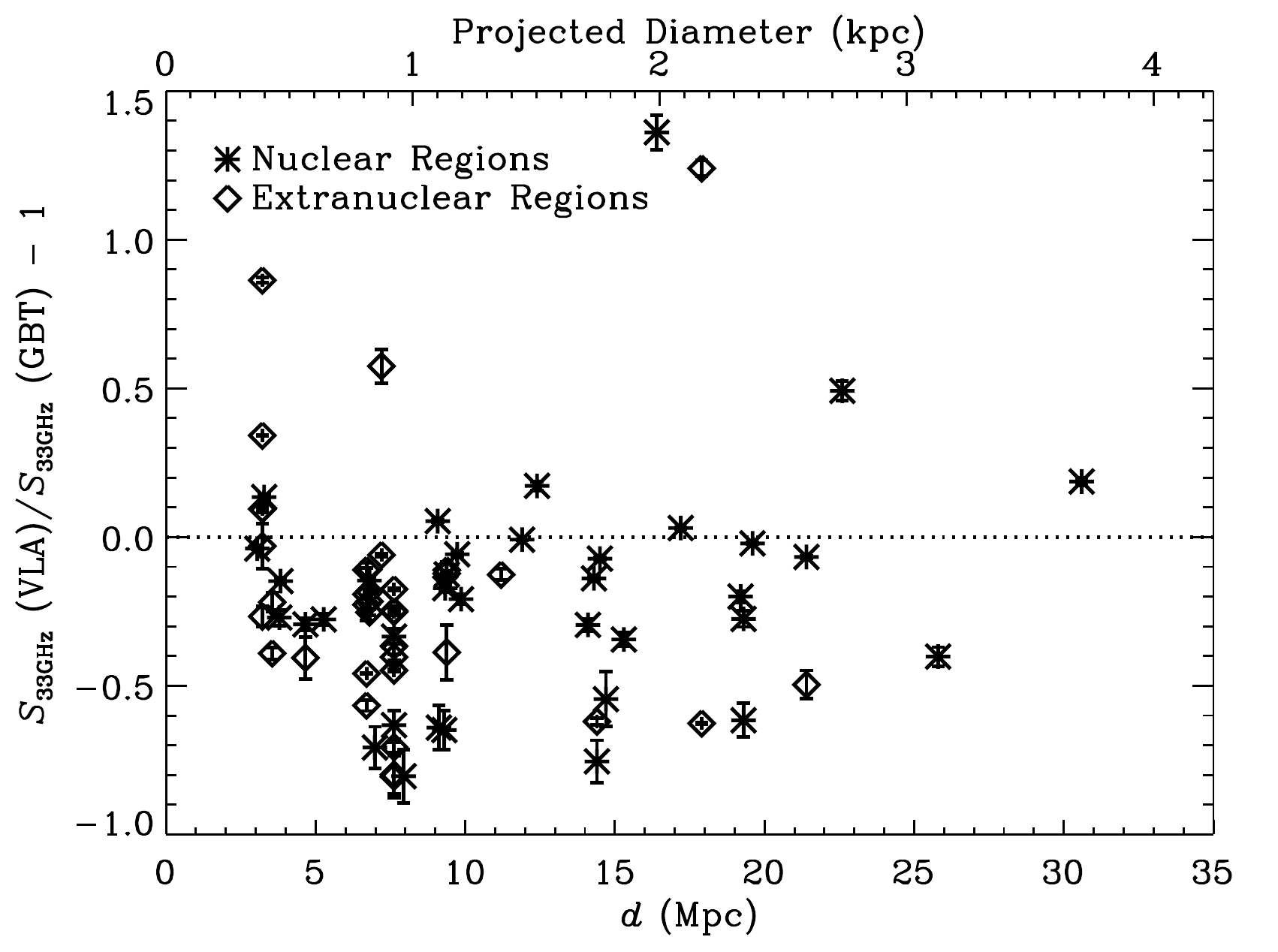}{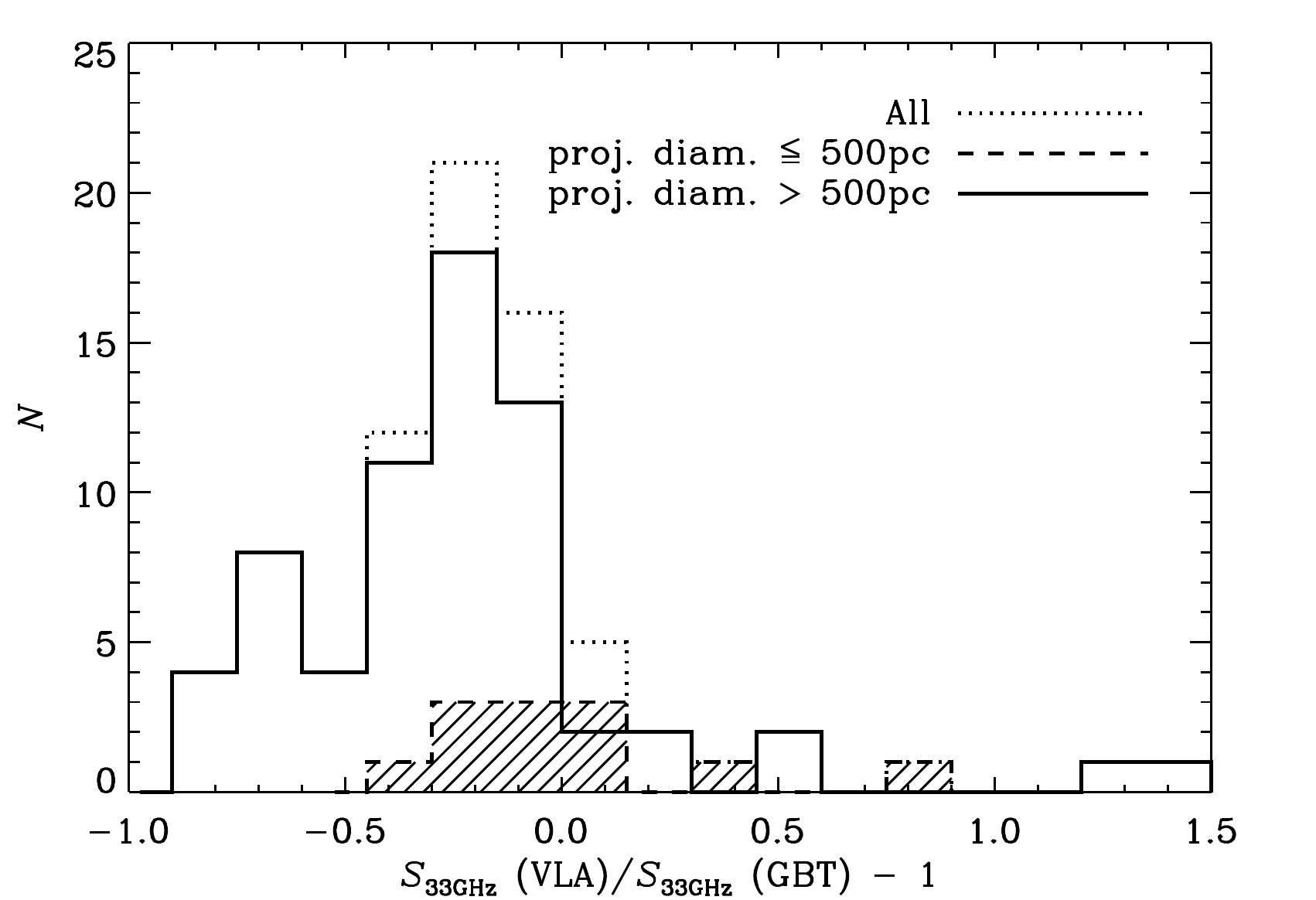}
\caption{
{\it Left:} The relative difference between the VLA and GBT measured 33\,GHz flux densities plotted against distance for sources detected at the 5$\sigma$ significance level in both datasets.  
The upper abscissa identifies the size of the projected diameter of the 25\arcsec~GBT beam.  
For the VLA flux densities were measured by multiplying the VLA image by an elliptical Gaussian to simulate the GBT observations (see \S\ref{sec:gbtcomp}).  
NGC\,4579, which hosts an AGN, is the data point for which the 33\,GHz VLA flux density is more than a factor of 2 larger than the corresponding 33\,GHz GBT flux density.  
{\it Right:} A histogram of the relative difference between the VLA and GBT measured 33\,GHz flux densities for sources detected at the 5$\sigma$ significance level in both datasets using bins of 0.15 (dotted line).  
Individual histograms of those sources for which the projected diameter of the 25\arcsec~GBT beam is larger (solid line) or smaller (dashed line/hatch filled) than $\approx 500$\,pc are also shown.  
What is clearly evident is that the VLA flux densities are systematically lower than what was recovered by the GBT.  
The median 33\,GHz VLA-to-GBT flux density ratio is $0.78\pm0.04$ with median absolute deviation of 0.27.  
For the 12 sources in which the 25\arcsec~GBT beam projects to a linear diameter of $\lesssim500$\,pc, the median 33\,GHz VLA-to-GBT flux density ratio is  $0.97\pm0.10$ with a median absolute deviation of 0.28, suggesting that this difference between the GBT and VLA flux densities likely arises from diffuse non-thermal synchrotron emission associated with CR electrons as they propagate away from their birth sites in supernova remnants near H{\sc ii} regions.   
\label{fig:gbtcomp}}
\end{figure*}

\subsection{Comparison with Single Dish \label{sec:gbtcomp}}
As a first test to see how much emission might be resolved out of these snapshot-like 33\,GHz images, we perform a comparison between the emission recovered in the interferometric images with the photometry obtained with the GBT given in \citet{ejm12b}.  
To do this, we multiply the VLA image by an elliptical Gaussian of peak unity, having major/minor axes and position angles based on the interferometric synthesized beams such that the convolution of the two results in a circular Gaussian beam with a FWHM of 25\arcsec, to match the typical beam size of the GBT at 33\,GHz.  
Since we assume a perfect Gaussian and do not account for additional emission arising from sidelobes in the actual GBT beam, these simulated observations will in most cases only provide a lower limit compared to what was measured by the GBT.  
However, we assume that this is likely a small (few percent) effect given that the sidelobes from the GBT measurements, when measurable, had an amplitude that is  2\% of the beam peak, on average \citep{ejm12b}.  

We compare these measured flux densities against what was measured by the GBT as a function of galaxy distance in the left panel of Figure \ref{fig:gbtcomp} for sources detected at $>5\sigma$ in both datasets.  
In the right panel of Figure \ref{fig:gbtcomp}, we plot the histogram of these sources using bins of 0.15 and highlight sources for which the 25\arcsec~GBT beam projects to a linear diameter of $\lesssim500$\,pc. 
What we find is that the VLA is typically missing $\approx$20\% of the total flux density recovered by the GBT.  
The median 33\,GHz VLA-to-GBT flux density ratio is  $0.78\pm0.04$ with a median absolute deviation of 0.27.    
The most likely reason for this discrepancy between the VLA and GBT photometry is that the GBT beam is picking up diffuse emission extended on scales greater than the largest-angular scale that these VLA 33\,GHz data are sensitive to (i.e., $\gtrsim24\arcsec$).  
However, we do not expect this to affect our aperture photometry results since we are only integrating on selected bright regions on the scale of a few arcseconds, where contributions from large scale diffuse emission on scales $\gtrsim 24\arcsec$ should be negligible.  

Furthermore, on such scales the bulk of the emission being resolved out by our 33\,GHz interferometric observations is likely diffuse non-thermal synchrotron emission associated with CR electrons as they propagate away from their birth sites in supernova remnants near H{\sc ii} regions.   
For example, the 12 sources in which the 25\arcsec~GBT beam projects to a linear diameter of $\lesssim500$\,pc, the median 33\,GHz VLA-to-GBT flux density ratio is  $0.97\pm0.10$ with a median absolute deviation of 0.28.   
Thus, given that the average thermal fraction at 33\,GHz reported by \citet{ejm12b} was 76\% for their entire sample (and $>$90\% on average for sources resolved on scales $\lesssim$500\,pc), this suggests that on the $\approx 30-300$\,pc scales of these VLA observations the 33\,GHz thermal fractions are most likely $\gtrsim 90$\%.  
Consequently, the 80\% thermal fraction of the GBT analysis is completely consistent with the measured 33\,GHz VLA-to-GBT flux density ratio if all compact emission is powered by free-free radiation while the non-thermal component is completely diffuse.  
We note that there is a minority of sources having values above unity, for which the VLA appears to be recovering more emission than the GBT.  
These occurrences most likely arise due to sources hosting a variable AGN (e.g., NGC\,4579 for which the 33\,GHz VLA flux density is more than a factor of 2 larger than the corresponding 33\,GHz GBT flux density) or situations where the GBT reference beam used for sky subtraction by nodding 1\farcm3 away from the source position landed on bright regions of the galaxies \citep[e.g., NGC\,3938\,Enuc.\,1; see][]{ejm12b}.

\subsection{33\,GHz and H$\alpha$ Morphologies \label{sec:hacomp}}

At the 2\arcsec~($\approx 30-300$\,pc) scales probed by our 33\,GHz observations, we are primarily sensitive to compact emission from individual star-forming complexes and galaxy nuclei. 
As a visual demonstration of this, we compared the 33\,GHz morphologies of our targets with their H$\alpha$ and 24\,$\mu$m morphologies. 
For highest resolution, the $\approx$2\arcsec~beam radio images and $\approx1-2$\arcsec~seeing limited H$\alpha$ images were compared at their native resolutions. 
To match the {\it Spitzer} PSF for the 33\,GHz/24\,$\mu$m comparison, we smoothed the 33\,GHz images to a 7\arcsec~circular beam with the CASA task {\sc imsmooth}. 
No astrometric alignment was necessary for the 24\,$\mu$m images, since the {\it Spitzer} astrometry was a near perfect match to the VLA astrometry at 7\arcsec~resolution. 

Figure~\ref{fig:imgs} shows H$\alpha$ and 24\,$\mu$m brightness contours overlaid on the 33\,GHz images. 
From a visual comparison, we find that at 7\arcsec~($\sim0.1 - 1$\,kpc) resolution, all but one strongly-detected ($\geq 5\sigma$) 33\,GHz source has a 24\,$\mu$m counterpart with a nearly identical morphology.
%From a visual comparison, it is clear that at 7\arcsec~resolution, every strongly detected ($\gtrsim 5\sigma$) 33\,GHz source has a 24\,$\mu$m counterpart with a nearly identical morphology. 
Such a tight morphological correlation is expected based on the well-known far-infrared (FIR)-radio correlation \citep{gxh85,de85}.
Studies of the resolved FIR-radio correlation \citep[e.g.,][]{ejm06a, ah06, fat07a, ejm08} find that lower frequency (synchrotron-dominated) radio emission is generally more spread out and diffuse than the corresponding dust emission associated with a single star-forming region as the result of CR electrons propagating significantly further than dust-heating photons.  
Since these 33\,GHz data are dominated by free-free emission rather than non-thermal synchrotron emission that traces propagating CR electrons, we expect this emission to remain more compact and closer to the skin of the H{\sc ii} regions where most of the warm dust emission is being powered.  
%(sentence about the FIR - radio correlation? something like This shows that the FIR - radio correlation (or is it the FIR-radio correlation with high frequency radio / 24um?) holds for both galaxy nuclei and extranuclear star forming complexes brighter than (threshold which corresponds to $SFR = 10^-3$\,Msun/yr) down to spatial scales of $\sim$ 90 - 900 pc). 
%Similarly, at a $1-2\arcsec$ resolution, the H$\alpha$/33\,GHz correspondence is similarly strong.  %(will do a thorough search for highly obscured HII regions & comment about that).  

\begin{figure}[t!]
\epsscale{1.2}
\plotone{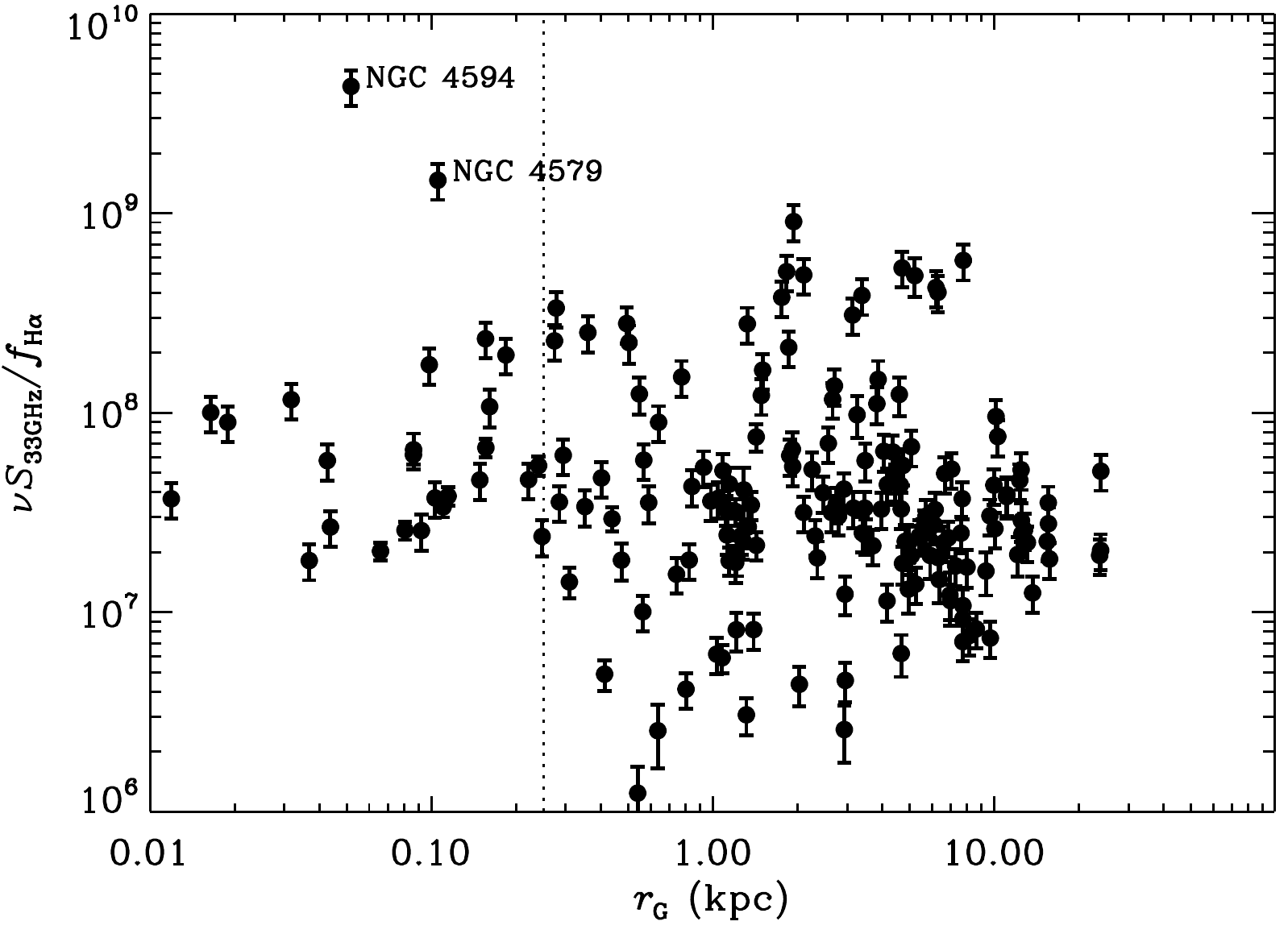}
\caption{The ratio of 33\,GHz flux to H$\alpha$ line flux plotted against galactocentric radius for all 162 sources having $\geq 3 \sigma$ detections at 33\,GHz and in H$\alpha$.  
The vertical line at $r_{\rm G} = 250$\,pc indicates the radius used to conservatively distinguish nuclear and extranuclear regions as some nuclear regions may be affected by AGN.  
While no obvious trend is seen, the median ratio does appear to be statistically larger within a central diameter of 500\,pc for all galaxies than the outer disks by a factor of $1.82\pm0.39$.  
We identify those sources that are clear outliers: NGC\,4594 and NGC\,4579, which are both known to harbor AGN that likely dominate the 33\,GHz continuum emission.  
\label{fig:fluxrat_rad}}
\end{figure}

At 2\arcsec~($\approx 30 - 300$\,pc) resolution, we find only four 33\,GHz sources that are plausibly associated with star-forming regions and do not have H$\alpha$ counterparts. 
Of these, two (NGC\,4631\,E and NGC\,4631\,F; see Figure \ref{fig:imgs}) are located in NGC\,4631, an edge-on spiral galaxy where dust lanes are likely strongly affecting the observed spatial distribution of H$\alpha$ emission. The first of the two remaining 33\,GHz/H$\alpha$ mismatches, NGC\,3627\,Enuc.\,1\,A, has a bright 33\,GHz peak that is morphologically distinct from any nearby H$\alpha$ structure and is offset from the nearest H$\alpha$ peak by $\approx$150\,pc. 
However, the 24\,$\mu$m peak pixel is co-located with the 33 GHz peak to better than $\approx$50\,pc. 
From this, we suspect that NGC\,3627\,Enuc.\,1\,A may be a highly extincted ($A_{\rm H\alpha} \gtrsim 5$\,mag) H{\sc ii} region. %(maybe something about follow up observations?) 
The final mismatch is NGC\,5194\,Enuc.\,11\,C, which is an unresolved radio peak located at the tip of a diffuse radio structure extending from the bright H{\sc ii} region NGC\,5194\,Enuc.\,11\,E.  
This source has neither an H$\alpha$ nor a 24\,$\mu$m counterpart, which rules out dust as an explanation for the mismatch. 
%{\bf Based on our 3 GHz and 15 GHz images which will be published in a forthcoming paper (? in prep), we measure a spectral index of XX for this source. (Sean, can you check the spectral index of this source which is at 13:29:49.5, +47.14.00.034 ?) 
%This suggests that XX (supernova remnants could have fairly flat spectral indices - there?s a Dave Green paper about this? or if it?s steep spectrum it might be a background source?). }

Our main result from this analysis is that $\approx$99\% of the 33\,GHz sources in our sample have morphologically similar counterparts in both the 24\,$\mu$m (on scales of a few hundred pc) and H$\alpha$ (on scales of $\sim$100\,pc) images. 
The striking morphological similarities between the three tracers suggest that for each of these regions, the H$\alpha$, 24\,$\mu$m and 33\,GHz emission are powered by the same source, namely massive star formation. 
The H$\alpha$ correspondence in particular suggests that the 33\,GHz emission is primarily powered by free-free emission. 
Another interesting implication of the 99\% matching between 33\,GHz (and 24\,$\mu$m) sources to H$\alpha$ sources is that this places a relatively strong limit on the number of deeply embedded bright star-forming regions in these galaxies.  
Using 24\,$\mu$m and H\,$\alpha$ observations  \citet[][]{mkmp07} report that $\approx$4\% of their sources are ``highly embedded" (i.e., $A_{\rm H\alpha} \gtrsim 3.3$\,mag) on $\approx$500\,pc scales for $\approx$1800 star-forming regions.  
Using that same criterion, we find that $\approx$10\% of our sources appear to be highly embedded (see \S\ref{sec:rad}).  
This is a slightly higher fraction than that reported by \citet[][]{mkmp07}, which may be due to sampling regions at finer spatial scales (i.e., $\approx$100\,pc compared to $\approx$500\,pc), or simply due to having much fewer sources in our analysis.  
%as pointed out previously by \citet[][]{mkmp07} who reported $\gtrsim$96\% association of 24\,$\mu$m and H\,$\alpha$ regions on $\approx$500\,pc scales for $\approx$1800 star-forming regions.  
If young clusters were buried in molecular clouds for a long period, we would expect to observe many 33\,GHz and 24\,$\mu$m sources without optical counterparts. 
Taking a typical H{\sc ii} region lifetime to be $\sim 5-10$\,Myr, our highly embedded fraction of $\approx$10\% suggests that, on average, an H{\sc ii} region remains embedded for $\lesssim1$\,Myr, consistent with multi-wavelength observations of young star-forming regions in a variety of extragalactic systems \citep[e.g.,][]{kej01,bcm11}.  
%(Johnson et al. 2011, ApJ, 559, 864, Whitmore et al. 2011, ApJ, 729, 78).}
%If young clusters were buried in molecular clouds for a long period, we would expect to observe many 33\,GHz and 24\,$\mu$m sources without optical counterparts.  
%{\bf Taking a typical H{\sc ii} region lifetime to be $\sim 5-10$\,Myr, our obscured fraction of $\approx$1\% suggests a region remains embedded for $\sim$0.5\,Myr and that it takes $\sim 2-3$\,Myr for the gas/dust to disperse, consistent with (Goddard et al. 2010, Bastian 2016 (review article)). }
%implies that a region would spend $\sim 5\times10^{5}$\,yr being obscured. 

\begin{figure}[t!]
\epsscale{1.25}
\plotone{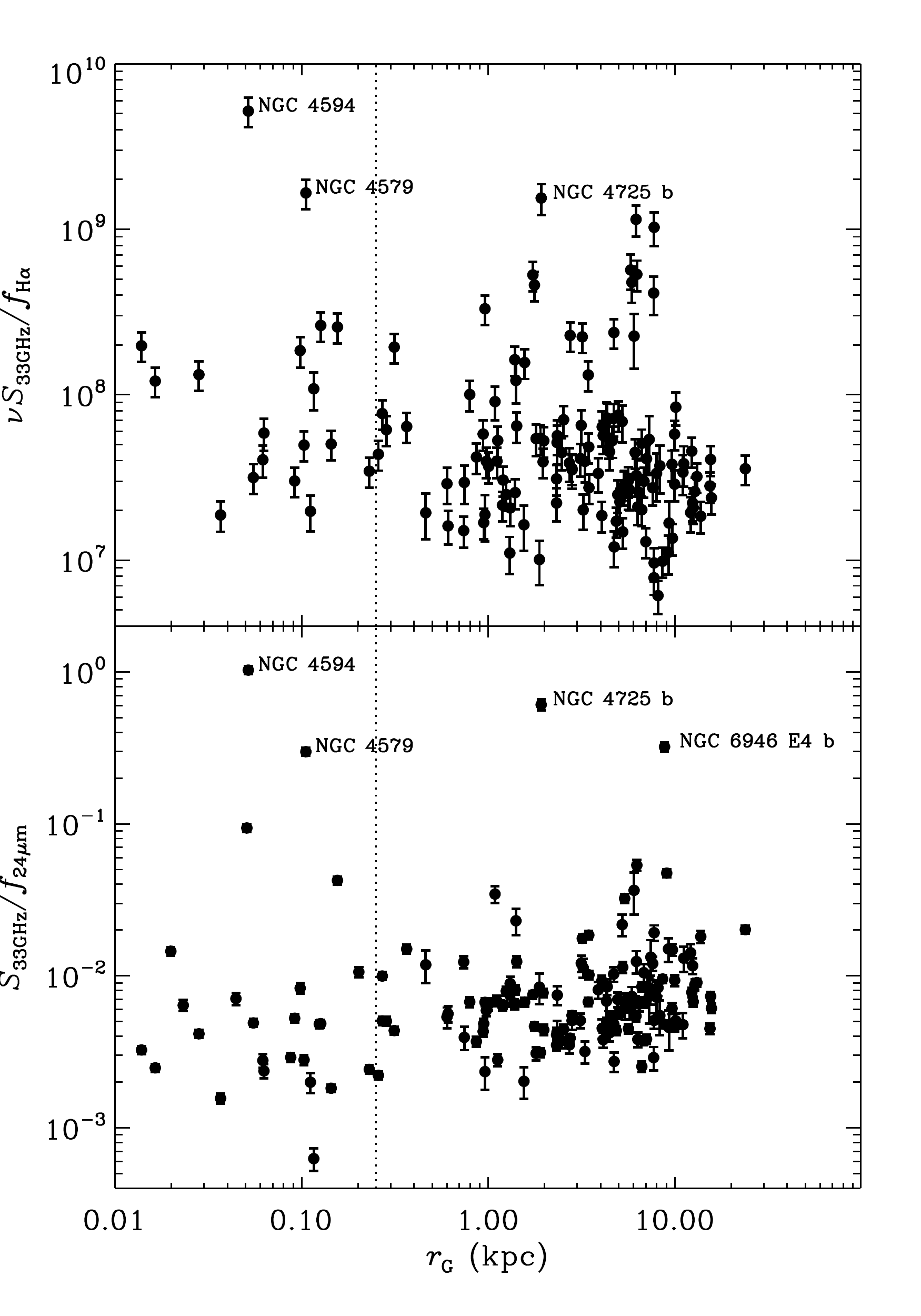}
%\plottwo{f4.pdf}{f5.pdf}
\caption{{\it Top:} The ratio of 33\,GHz flux to H$\alpha$ line flux plotted against galactocentric radius for all 144 sources having $\geq 3 \sigma$ detections at 33\,GHz and in H$\alpha$ after convolving both data set to 7\arcsec~resolution to match the resolution of the 24\,$\mu$m {\it Spitzer} data.    
The vertical line at $r_{\rm G} = 250$\,pc (in both panels) indicates the radius used to conservatively distinguish nuclear and extranuclear regions as some nuclear regions may be affected by AGN.    
Similar to what is plotted in Figure \ref{fig:fluxrat_rad} at higher resolution, no obvious trend is seen.  
However, the median ratio does appear to be larger within a galactocentric radius  $r_{\rm G} < 250$\,pc for all galaxies than the outer disks by a factor of $1.53\pm0.55$.  
{\it Bottom:}  The ratio of 33\,GHz to 24\,$\mu$m flux density plotted against galactocentric radius for all 160 sources having $\geq 3 \sigma$ detections at 33\,GHz and 24\,$\mu$m.  
Similar to the top panel, no obvious trend with galactocentric radius is seen.  
However, the median ratio does appear to be significantly smaller within a galactocentric radius $r_{\rm G} < 250$\,pc for all galaxies compared to the outer disks by a factor of $0.45\pm0.08$.  
In both panels we identify those sources that are clear outliers: NGC\,4594 and NGC\,4579, which are both known to harbor AGN; NGC\,6946\,Enuc.4\,B, which is a known AME detection \citep{ejm10, as10, bh14}; and NGC\,4725\,B, which may be a background AGN or another AME detection and warrants further investigation.  
\label{fig:fluxrat_rad_tap}}
\end{figure}

\subsection{Radial Trends \label{sec:rad}}
In Figure \ref{fig:fluxrat_rad} we investigate if there are any trends in the ratio of the 33\,GHz flux to H$\alpha$ line flux as a function of galactocentric radius.  
We distinguish nuclear from extranuclear sources as having a galactocentric radius $r_{\rm G} < 250$\,pc since in some cases a fraction of the nuclear 33\,GHz emission may be powered by a central AGN (see Figure \ref{fig:fluxrat_rad}).  
For the 162 extranuclear sources detected at $> 3\sigma$ significance at both 33\,GHz and in H$\alpha$, we calculate star formation rates following the equations given in \citet{ejm11b,ejm12b}.  
As discussed in \S\ref{sec:gbtcomp}, given that these 33\,GHz data are able to resolve star-forming regions within each galaxy on $\approx$100\,pc scales combined with the results of  \citet{ejm12b,ejm15}, we assume a 33\,GHz thermal fraction of $\approx$90\% {when calculating star formation rates with Equation~11 in \citet{ejm11b}.  
Using the ratio of the optically-thin 33\,GHz to uncorrected H$\alpha$ star formation rates, we calculate a median extinction value on $30-300$\,pc scales of $A_{\rm H\alpha} \approx 1.26\pm0.09$\,mag, similar to the value of 1.4\,mag reported by \citet{mkmp07} when comparing 24\,$\mu$m and H\,$\alpha$ photometry on 500\,pc scales for nearly 1800 star-forming regions within a sample of 38 nearby galaxies.  
The associated median absolute deviation is 0.87\,mag.  
We believe that the rather large scatter here is driven by the corresponding large (20\%) calibration uncertainty associated with the difficulties in H$\alpha$ narrowband imaging.  

A strong trend in 33\,GHz flux to H$\alpha$ line flux with galactocentric radius is not observed, however the median ratio does appear to be statistically larger within the central 500\,pc diameter for all galaxies compared to the outer disks by a factor of $1.82\pm0.39$. 
Furthermore, a two-sided Kolmogorov-Smirnov (KS) test yields a probability of only $\approx$1.4\% that both sets of ratios are drawn from the same distribution.  
With only the 33\,GHz and H$\alpha$ data alone, it is unclear if this result is primarily due to a higher amount of non-thermal emission contributing to the 33\,GHz flux density or a larger amount of extinction attenuating the H$\alpha$ emission within a galactocentric radius $r_{\rm G} < 250$\,pc for these galaxies.  
It is worth noting that there are studies in the literature showing that thermal fractions of circumnuclear star-forming regions are indeed lower relative to those in the outer disks of galaxies \citep[e.g.,][]{rck89, ejm11b}, which argues that additional non-thermal emission likely plays a role, although there is significant scatter among sources \citep[e.g.,][]{ejm12b}.  

To attempt to break this degeneracy, we again plot the ratio of the 33\,GHz flux to H$\alpha$ line flux as a function of galactocentric radius in the top panel of Figure \ref{fig:fluxrat_rad_tap} as well as the ratio of the 33\,GHz to 24\,$\mu$m flux density ratio as a function of galactocentric radius in the bottom panel, all at matched resolution.  
Of the 179 discrete regions used for aperture photometry in the convolved maps, there are a total of 144 and 160 sources detected at $\geq 3 \sigma$ at 33\,GHz and H$\alpha$ and 24\,$\mu$m, respectively.  
In both panels we identify those sources with ratios that are clear outliers.  
These include NGC\,4594 and NGC\,4579, which are both known to harbor AGN, NGC\,6946\,Enuc.4\,B, which is a known AME detection \citep{ejm10, as10, bh14}.  
The final source, NGC\,4725\,B has a spectrum that rises between 15 and 33\,GHz based on data to be published in a forthcoming paper.  % (D. Dong et al., 2017, in preparation).  
This may be indicative of another AME detection, but requires further investigation to see if this is indeed the case, or perhaps a background AGN peaking at $\gtrsim33$\,GHz.  
%based on our 3\,GHz and 15\,GHz images that will be published in a forthcoming paper (? in prep), we measure a spectral index of XX for this source.  which may be another AME detection and requires further investigation.  

Similar to what is found in Figure \ref{fig:fluxrat_rad} at higher resolution, the median ratio of 33\,GHz flux to H$\alpha$ line flux does appear to be larger within a galactocentric radius $r_{\rm G} < 250$\,pc for all galaxies relative to the outer-disk regions by a factor of $1.53 \pm 0.55$.  
A two-sided KS test yields a probability of $\approx$5\% that both sets of ratios are drawn from the same distribution, which is less significant than the value measured at higher angular resolution above (i.e., $\approx$1.4\%).  
Assuming that the 33\,GHz and 24\,$\mu$m emission are both tracing current star formation unbiased by dust, any increase in this ratio among the nuclear versus the extranculear regions would suggest that the difference in the 33\,GHz flux and H$\alpha$ line flux ratios are in fact due to an additional emission component powering the 33\,GHz emission (i.e., additional non-thermal emission).  
While we again find no obvious trend between the ratio of the 33\,GHz to 24\,$\mu$m flux densities versus galactocentric radius, the median ratio actually appears significantly \textit{smaller} within a galactocentric radius $r_{\rm G} < 250$\,pc for all galaxies compared to the outer disks by a factor of $0.45 \pm 0.08$.    
A two-sided KS test in this case yields a probability of $\ll$1\% that both sets of ratios are drawn from the same distribution.  
Consequently, there appears to be a larger amount of warm dust emission per unit star-formation activity compared to 33\,GHz emission within the central 500\,pc diameter for the sample galaxies, consistent with far-infrared studies of nearby galaxies that find dust tends to be warmer in the centers of galaxies \citep[e.g., ][]{fat07b,bg12,gjb15}.  

Such a situation may arise if the circumnuclear regions of these galaxies have undergone an extended star formation history in which star formation that has taken place over a longer period of time, resulting in an accumulation of $\gtrsim$3\,Myr dust-heating stars in addition to any very old bulge stars
 that boost the 24\,$\mu$m flux density relative to the extranuclear regions.  
This is largely opposite to what we would expect if there was an additional component of non-thermal emission powering the 33\,GHz in the central regions of these galaxies, unless the excess dust heating at 24\,$\mu$m far exceeds any additional non-thermal emission contribution at 33\,GHz.  
So, while this result alone suggests that the larger ratio of 33\,GHz flux to H$\alpha$ line flux found in the central regions of these galaxies may primarily arise from increased extinction, more detailed radio spectral fitting to obtain reliable thermal fractions is needed to help to confirm the dominant physical process driving the observed trend.

\section{Conclusions}
We have presented 33\,GHz interferometric imaging taken with the VLA for 112 fields (50 nuclei and 62 extranuclear H{\sc ii} regions) observed as part of the SFRS.  
These $\approx 2\arcsec$ resolution images are compared to archival H$\alpha$ and 24\,$\mu$m imaging.  
Our conclusions can be summarized as follows:

\begin{itemize}

\item{A comparison with GBT single-dish 33\,GHz observations indicates that the interferometric VLA observations recover $78\pm4\%$ of the total flux density over 25\arcsec~regions ($\approx$kpc-scales) among all fields on average indicating that, on the $\la 300$\,pc scales sampled by our VLA observations, missing emission from the lack of short spacings is not significant.  
On $\approx$kpc scales, the bulk of the emission being resolved out by our 33\,GHz interferometric observations is most likely diffuse non-thermal synchrotron emission associated with CR electrons as they propagate away from their birth sites in supernova remnants near H{\sc ii} regions.   
Consequently, on the $\approx 30 - 300$\,pc scales sampled by our VLA observations the observed 33\,GHz emission is primarily powered by free-free emission from discrete H{\sc ii} regions making it an excellent tracer of massive star formation.  
}

\item{A morphological comparison between the 33\,GHz radio, H$\alpha$ nebular line, and 24\,$\mu$m warm dust emission shows remarkably tight similarities in their distributions, suggesting that each of these emission components are indeed powered by a common source expected to be massive star-forming regions, and against suggests that the 33\,GHz emission is dominated by free-free emission.   
}

\item{Of the 225 discrete regions used for aperture photometry, 162 are detected at $> 3\sigma$ significance at both 33\,GHz and in H$\alpha$ and are conservatively considered to be extranuclear and star forming by having galactocentric radii $r_{\rm G} \geq 250$\,pc.  
By assuming a typical 33\,GHz thermal fraction of 90\%, we use this ratio of the optically-thin 33\,GHz to uncorrected H$\alpha$ star formation rates to calculate a median extinction value on $30-300$\,pc scales of $A_{\rm H\alpha} \approx 1.26\pm0.09$\,mag with an associated median absolute deviation of 0.87\,mag among these star-forming regions. }

\item{We find that $\approx$99\% of 33\,GHz sources in our sample have morphologically similar counterparts in both the 24\,$\mu$m (on scales of a few hundred pc) and H$\alpha$ (on scales of $\sim$100\,pc) images suggesting that each is powered by massive star formation. 
The H$\alpha$ correspondence in particular suggests that the 33\,GHz emission is primarily powered by free-free emission. 
This result additionally puts a limit on the number of deeply embedded bright star-forming regions in these galaxies given that if young clusters were buried in molecular clouds for a long period, we would expect to observe many 33\,GHz and 24\,$\mu$m sources without optical counterparts. 
Our ``highly embedded" (i.e., $A_{\rm H\alpha} \gtrsim 3.3$\,mag) fraction of $\approx$10\% suggests that, on average, H{\sc ii} regions remain embedded for $\lesssim 1$\,Myr.  
}

\item{We find that the median 33\,GHz flux to H$\alpha$ line flux ratio to be statistically larger within a galactocentric radius $r_{\rm G} < 250$\,pc for all galaxies relative to the outer-disk regions by a factor of $1.82\pm0.39$.  
We additionally find that the median 33\,GHz to 24\,$\mu$m ratio does appear to be statistically smaller in the central 500\,pc diameter for all galaxies compared to the outer-disk regions by a factor of $0.45\pm0.08$. 
The combination of these results suggests that the larger ratio of 33\,GHz flux to H$\alpha$ line flux found in the central regions may arise primarily by increased extinction, rather than an excess of non-thermal radio emission.
However, more detailed radio spectral fitting to obtain reliable thermal fractions is needed to help to confirm the dominant physical process driving this observed trend.  

}

\end{itemize}

\acknowledgements
We would like to thank the anonymous referee for very useful comments that helped to improve the content and presentation of this paper.
The National Radio Astronomy Observatory is a facility of the National Science Foundation operated under cooperative agreement by Associated Universities, Inc.
E.J.M. acknowledges the hospitality of the Aspen Center for Physics, which is supported by the National Science Foundation Grant No. PHY-1066293.  
This research made use of APLpy, an open-source plotting package for Python hosted at http://aplpy.github.com
If you are using APLpy 0.9.9 or later, please also include the acknowledgment for the Astropy package, since it is used heavily as a dependency.

\bibliography{aph.bbl}
%\bibliography{/Users/emurphy/libs/bibtexref/master_ref}

\setcounter{figure}{0}
\begin{figure*}[htb!]
\epsscale{1.0}
\plottwo{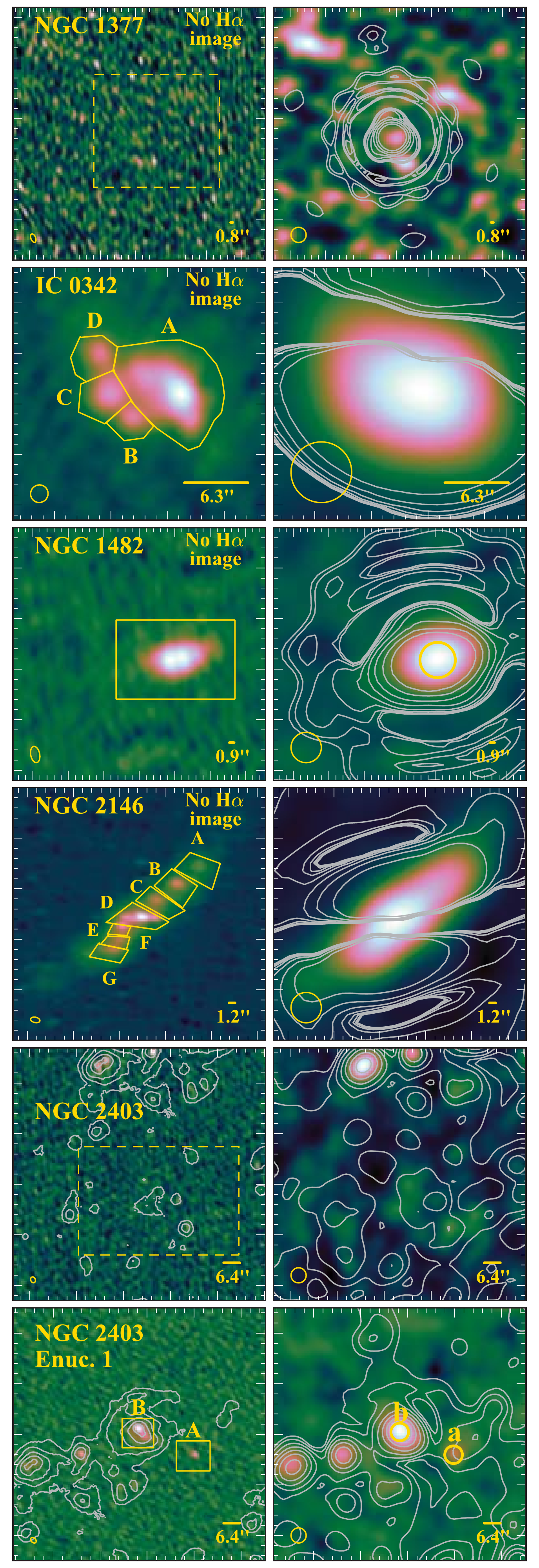}{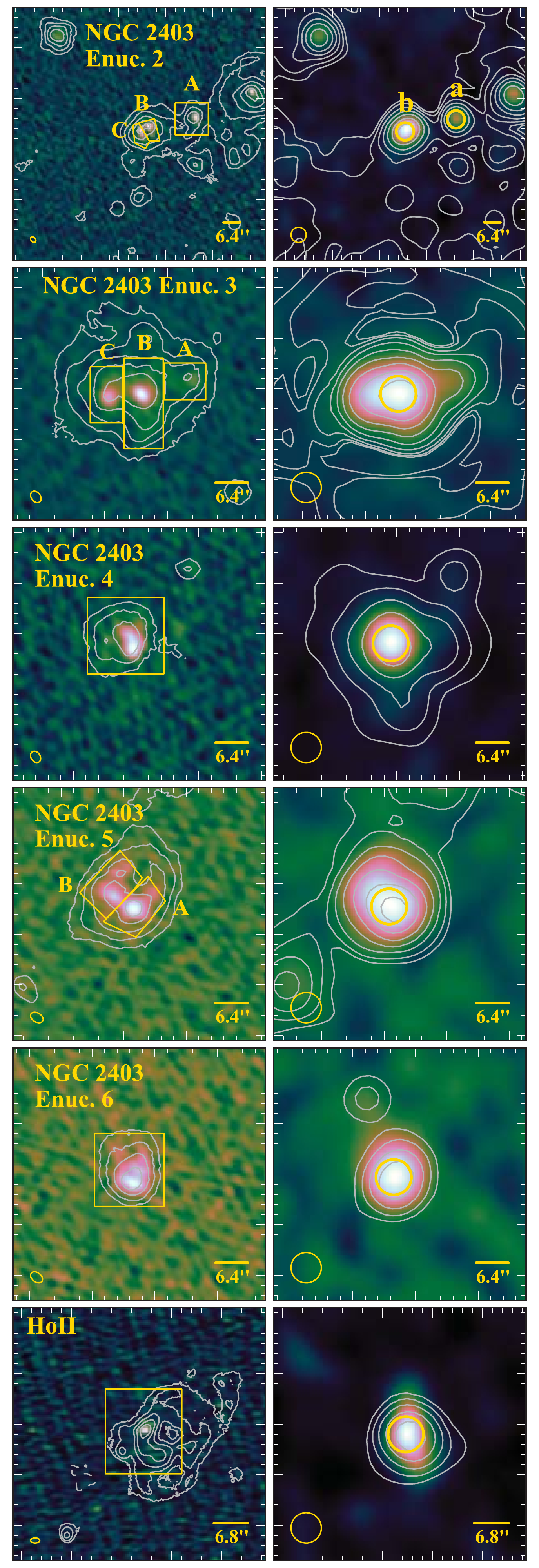}
\caption{ {\it Continued}
\label{fig:imgs}}
\end{figure*}

\setcounter{figure}{0}
\begin{figure*}[htb!]
\epsscale{1.0}
\plottwo{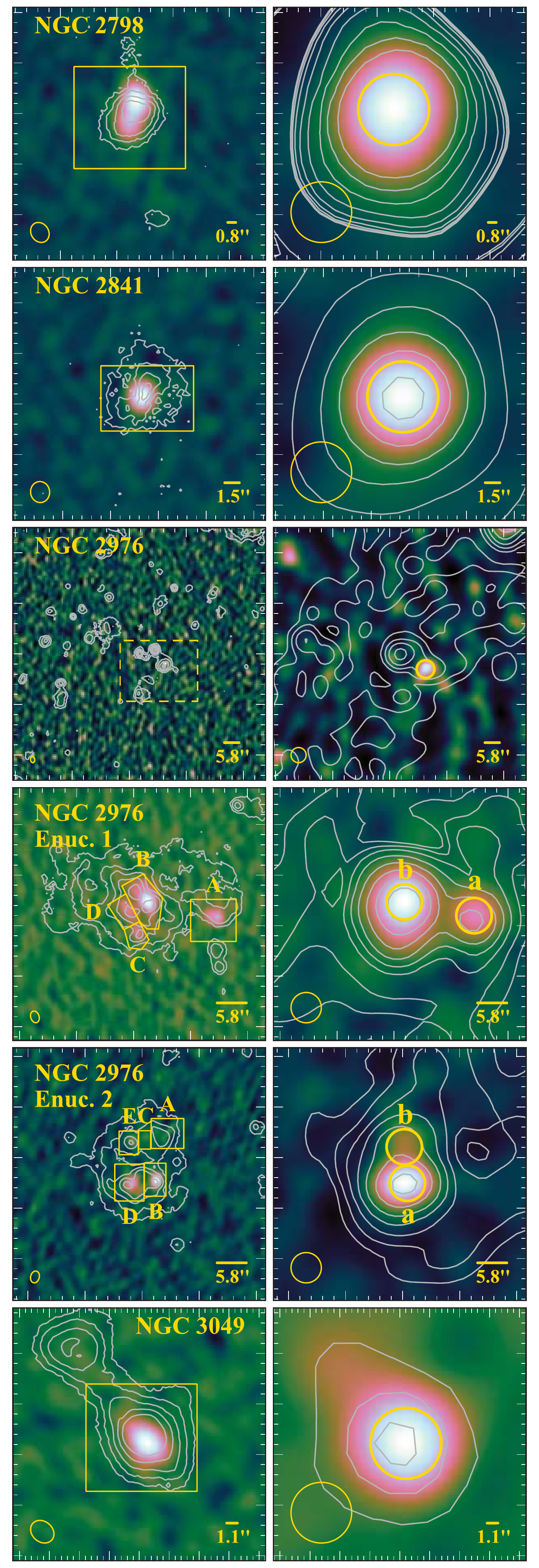}{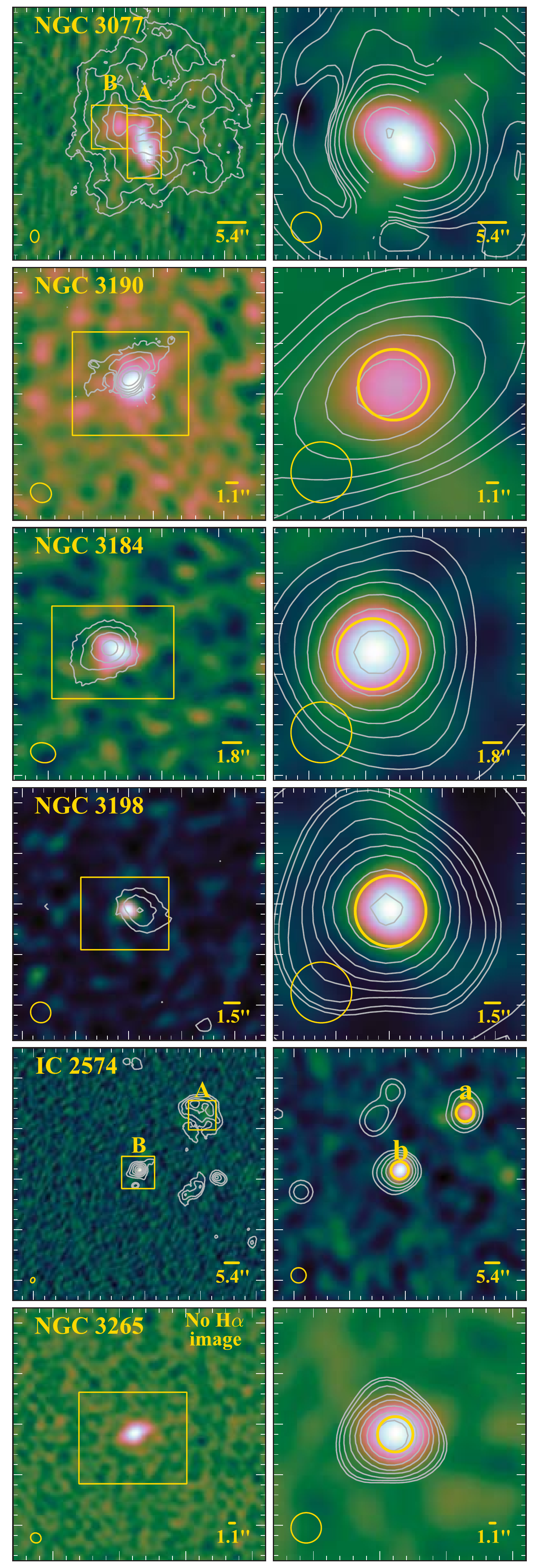}
\caption{  {\it Continued}
\label{fig:imgs}}
\end{figure*}

\setcounter{figure}{0}
\begin{figure*}[htb!]
\epsscale{1.0}
\plottwo{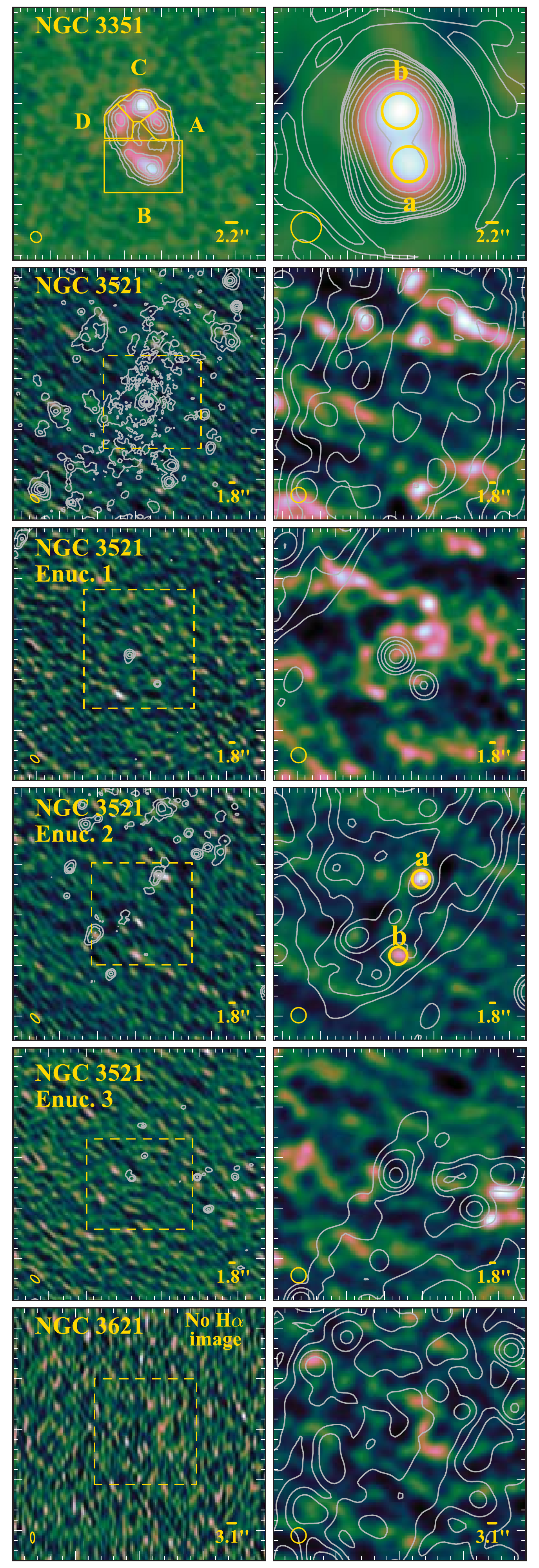}{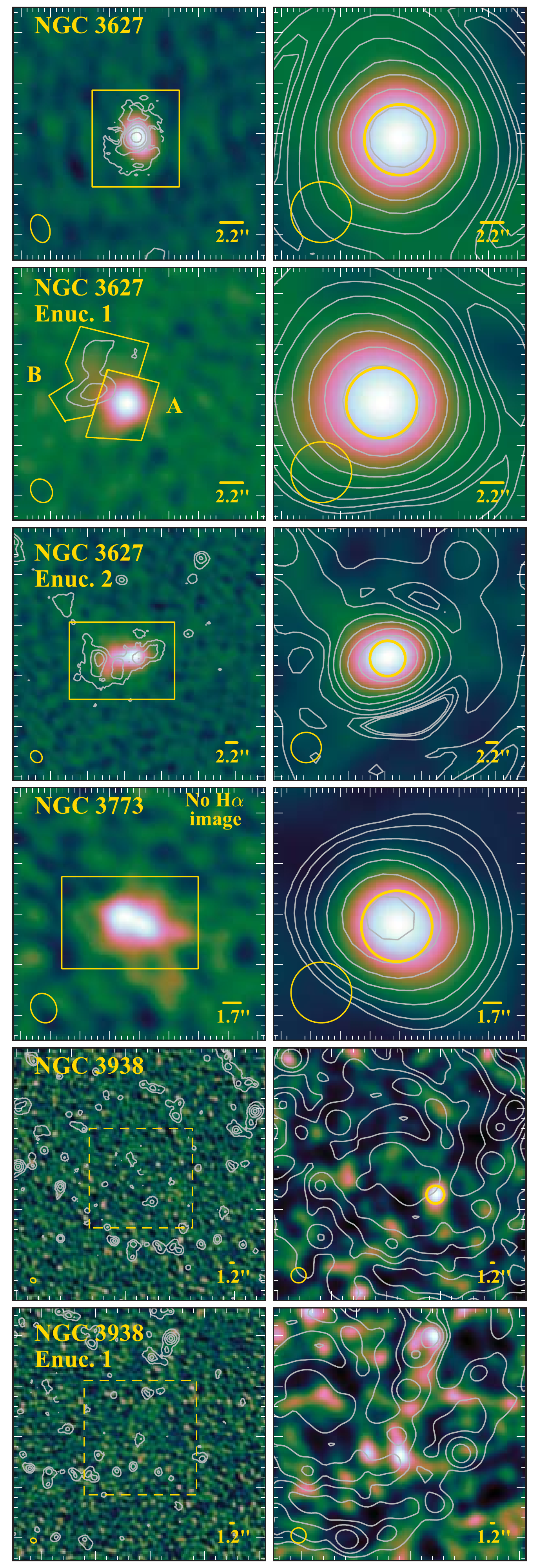}
\caption{ {\it Continued} 
\label{fig:imgs}}
\end{figure*}

\setcounter{figure}{0}
\begin{figure*}[htb!]
\epsscale{1.0}
\plottwo{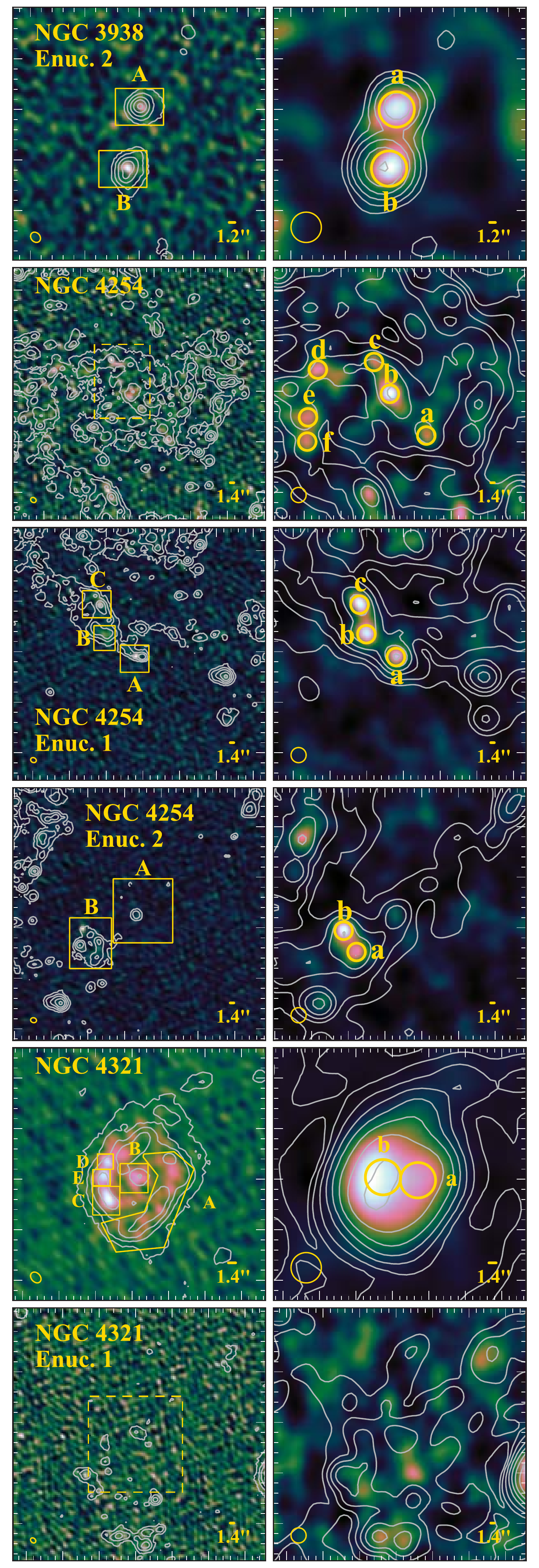}{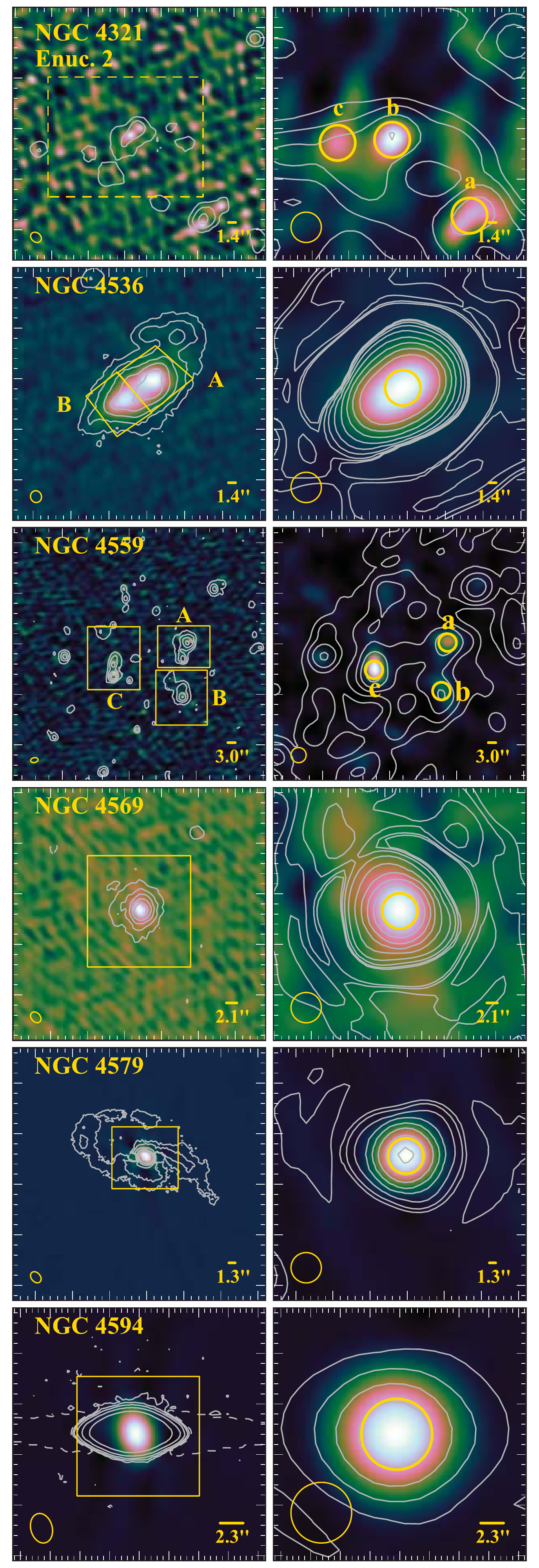}
\caption{ {\it Continued} 
\label{fig:imgs}}
\end{figure*}

\setcounter{figure}{0}
\begin{figure*}[htb!]
\epsscale{1.0}
\plottwo{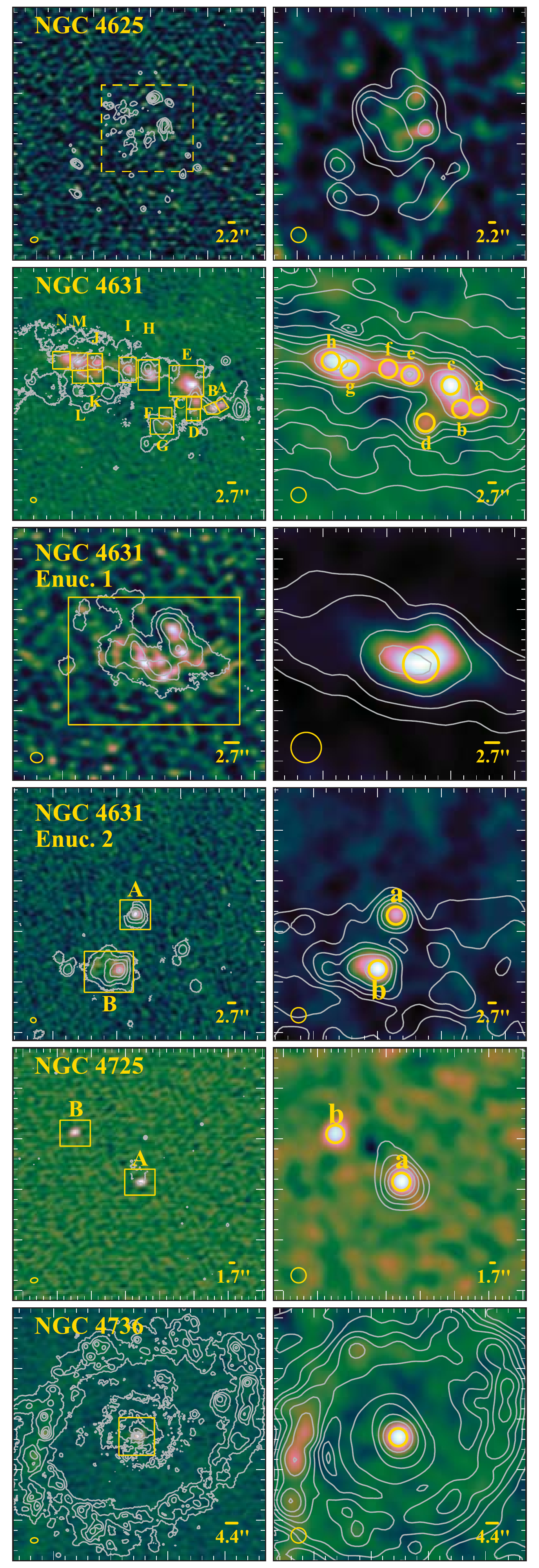}{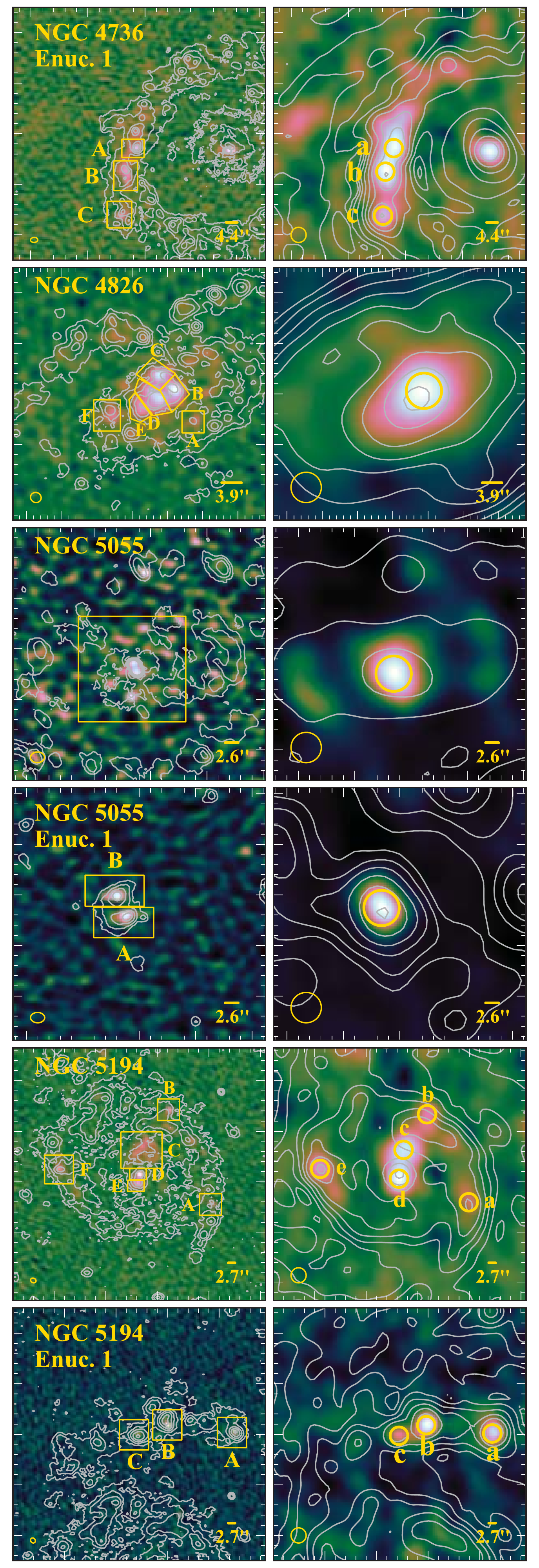}
\caption{ {\it Continued}
\label{fig:imgs}}
\end{figure*}

\setcounter{figure}{0}
\begin{figure*}[htb!]
\epsscale{1.0}
\plottwo{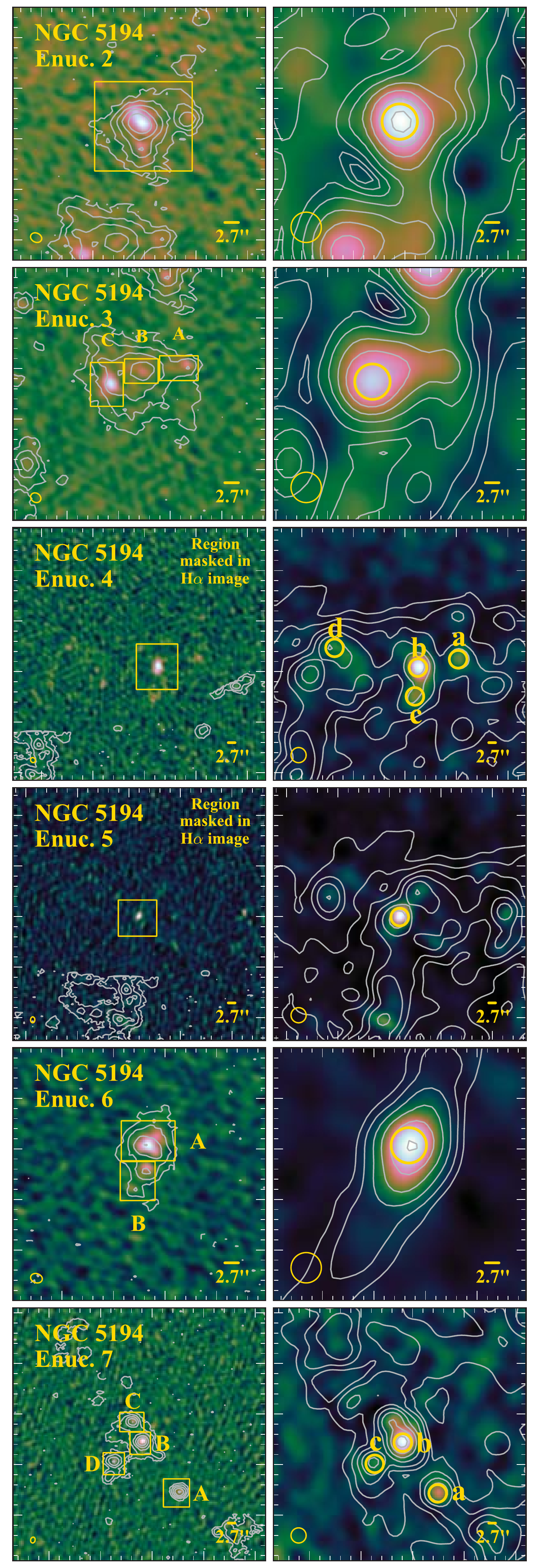}{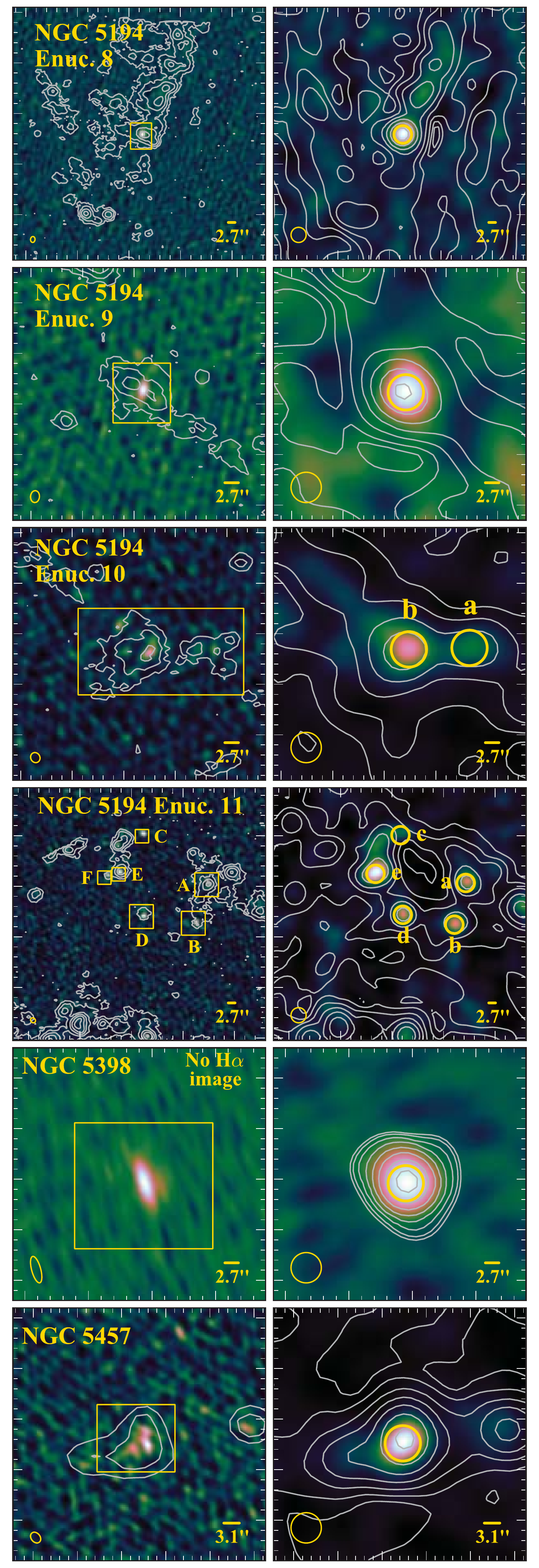}
\caption{ {\it Continued}
\label{fig:imgs}}
\end{figure*}

\setcounter{figure}{0}
\begin{figure*}[htb!]
\epsscale{1.0}
\plottwo{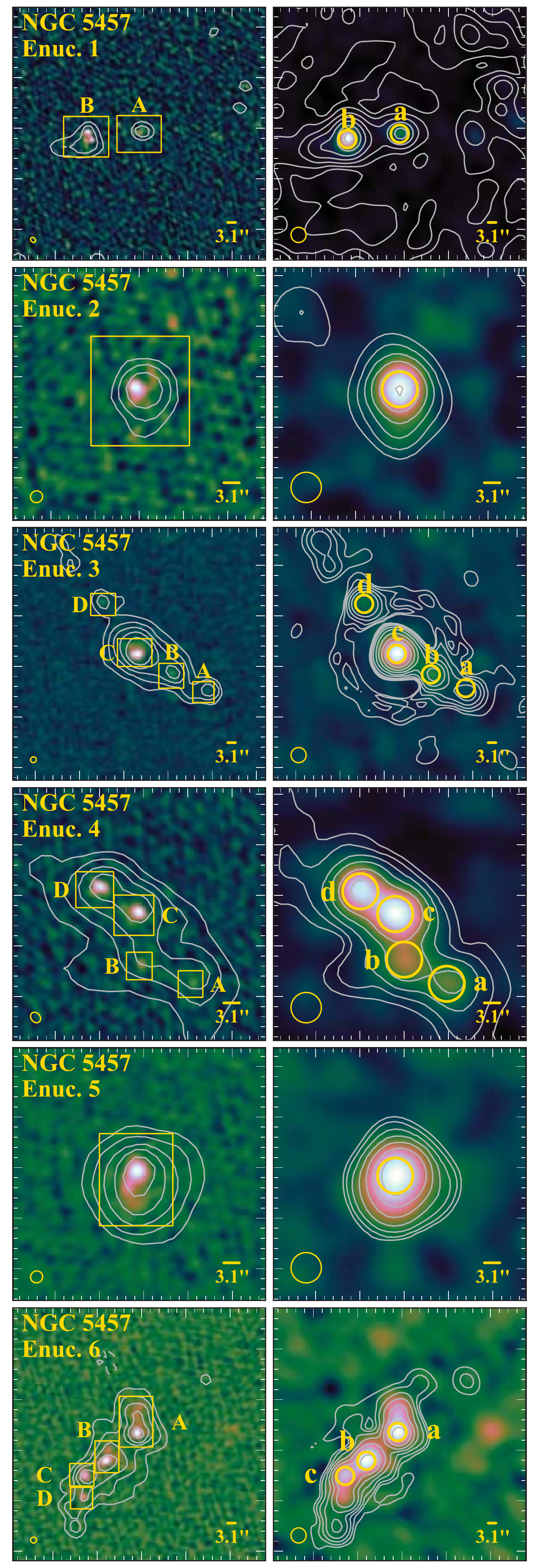}{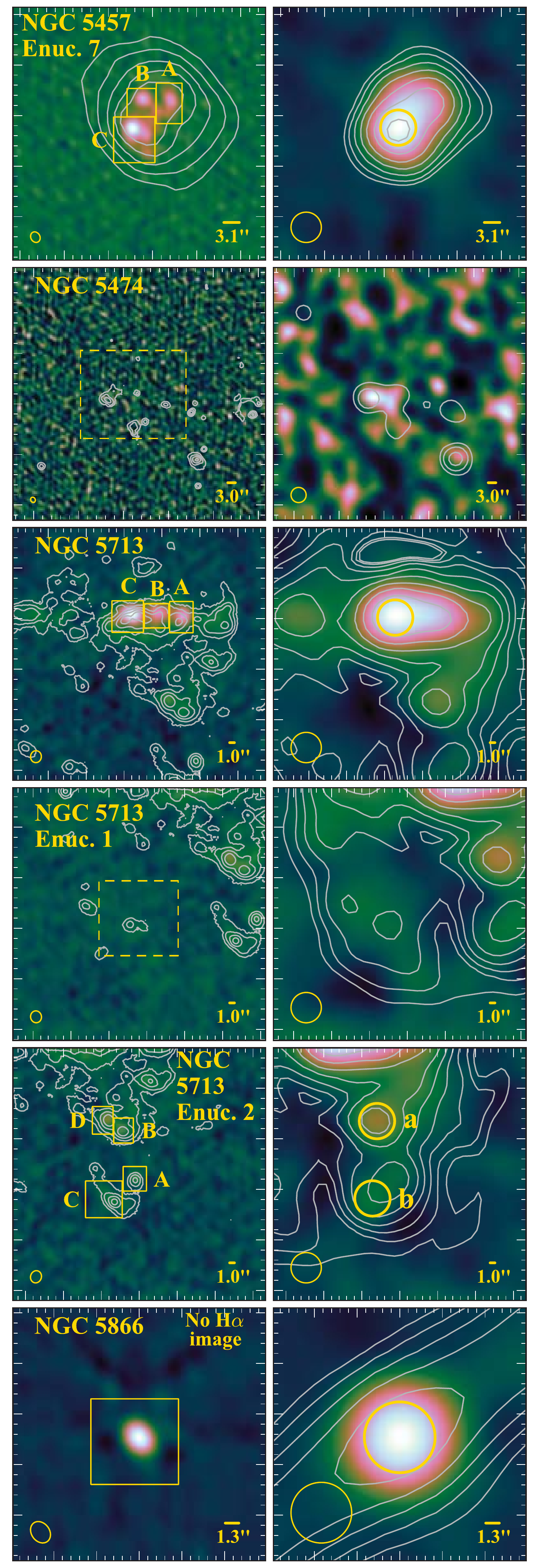}
\caption{  {\it Continued}
\label{fig:imgs}}
\end{figure*}

\setcounter{figure}{0}
\begin{figure*}[htb!]
\epsscale{1.0}
\plottwo{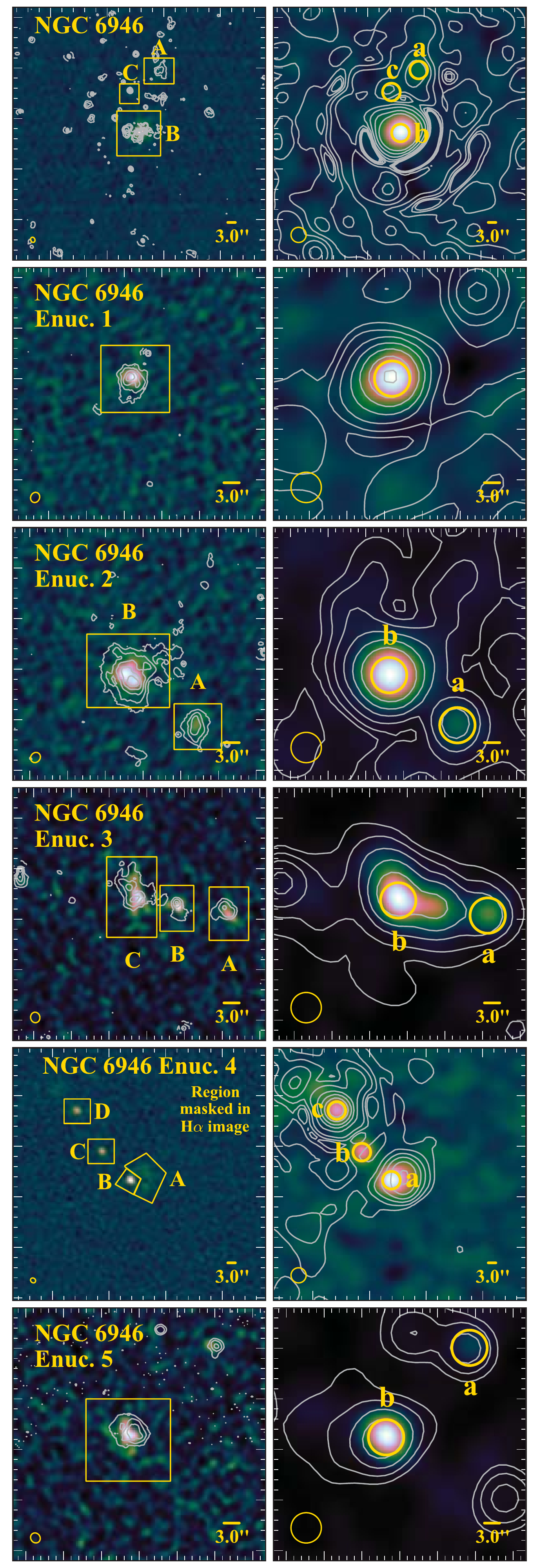}{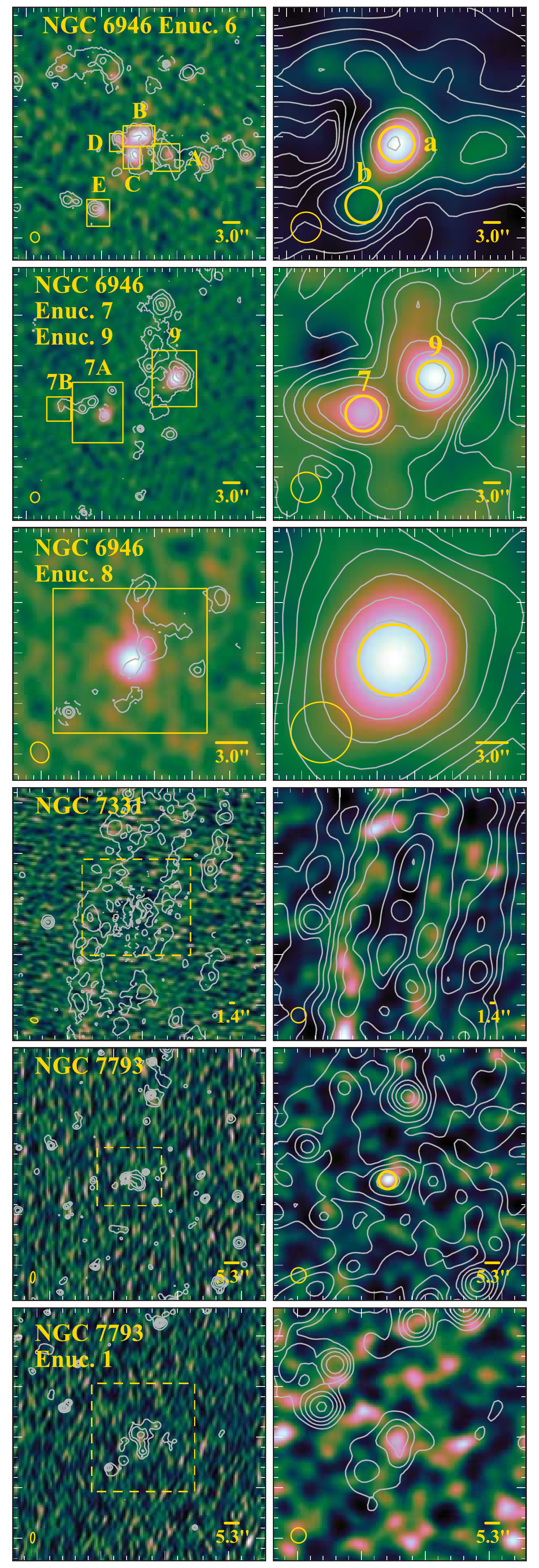}
\caption{ {\it Continued} 
\label{fig:imgs}}
\end{figure*}

\setcounter{figure}{0}
\begin{figure}[htb!]
\epsscale{1.}
\plotone{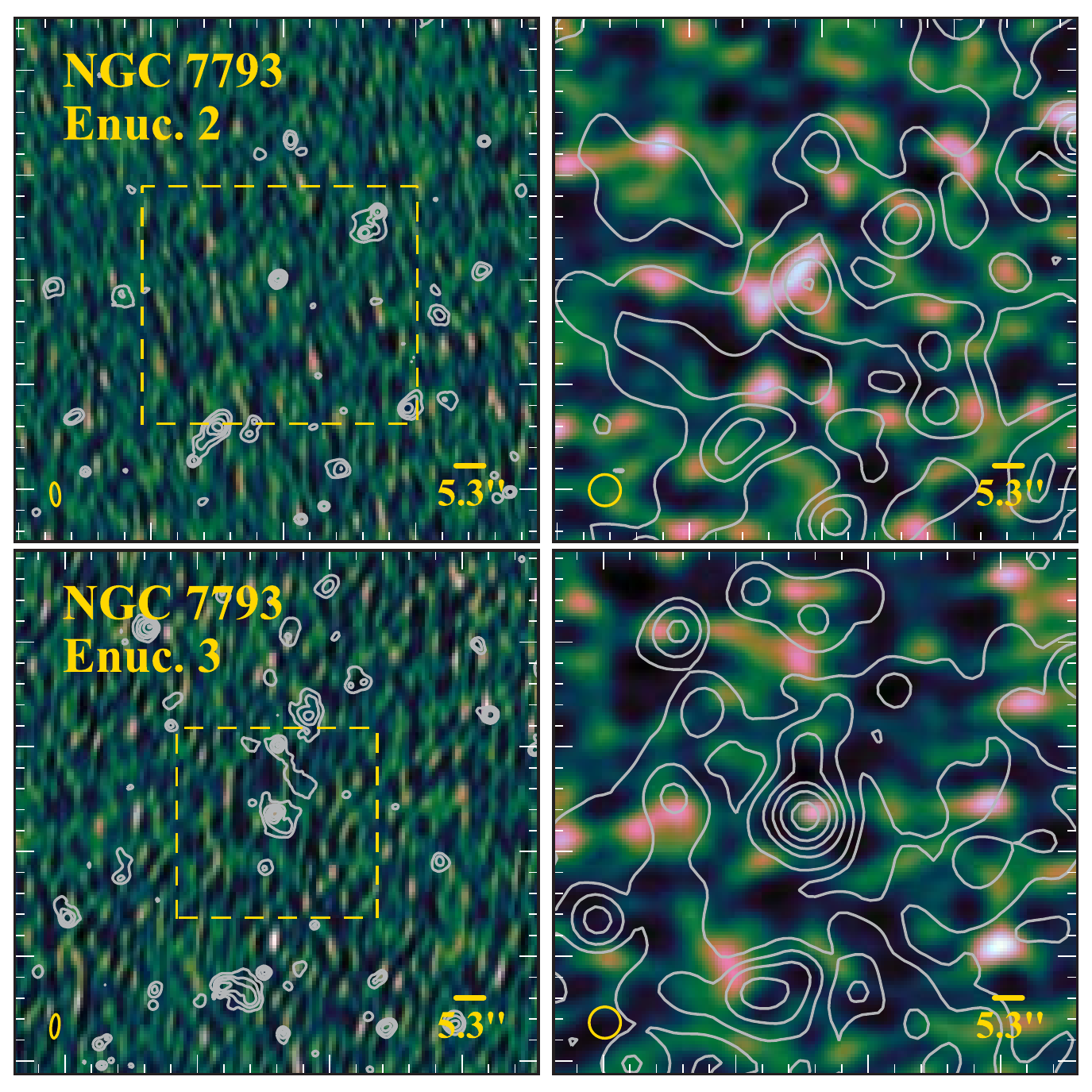}
\caption{ {\it Continued}
\label{fig:imgs}}
\end{figure}

\clearpage

\LongTables
\begin{deluxetable*}{l|cc|ccr}
\tablecaption{Source Photometry \label{tbl-5}}
\tablehead{
\colhead{Source ID} & \colhead{R.A.} & \colhead{Decl.} & \colhead{$S_\mathrm{33\,GHz}$} & \colhead{$f_\mathrm{H_{\alpha}}/10^{-13}$} & \colhead{$r_\mathrm{G}$}\\
\colhead{} & \colhead{(J2000)} & \colhead{(J2000)} & \colhead{(mJy)} & \colhead{(erg\,s$^{-1}$\,cm$^{-2}$)} & \colhead{(kpc)}
}
\startdata
                    NGC\,0337\,A   &           $00~59~50.00$  &           $-07~34~34.7$  & $  0.57\pm  0.02$  & $  3.40\pm  0.68$  & 0.823  \\
                    NGC\,0337\,B   &           $00~59~50.74$  &           $-07~34~58.7$  & $  1.79\pm  0.06$  & $  6.18\pm  1.24$  & 2.100  \\
                    NGC\,0337\,C   &           $00~59~51.77$  &           $-07~34~57.2$  & $  0.15\pm  0.01$  & $  1.29\pm  0.26$  & 2.946  \\
                    NGC\,0337\,D   &           $00~59~51.87$  &           $-07~34~36.2$  & $  1.01\pm  0.03$  & $  5.12\pm  1.02$  & 3.699  \\
             NGC\,0628\,Enuc.\,4   &           $01~36~35.71$  &           $+15~50~07.2$  & $  0.26\pm  0.01$  & $  1.15\pm  0.23$  & 7.611  \\
             NGC\,0628\,Enuc.\,2   &           $01~36~37.65$  &           $+15~45~07.2$  & $  0.48\pm  0.02$  & $  1.04\pm  0.21$  & 4.468  \\
             NGC\,0628\,Enuc.\,3   &           $01~36~38.92$  &           $+15~44~22.9$  & $  0.59\pm  0.03$  & $  2.14\pm  0.43$  & 5.715  \\
                       NGC\,0628   &           $01~36~43.26$  &           $+15~46~38.6$  & $  0.24\pm  0.02$  & $  3.26\pm  0.65$  & 1.210  \\
\enddata
\end{deluxetable*}

\begin{deluxetable*}{l|cc|cccr}
\tablecaption{Source Photometry at 7\arcsec~Resolution \label{tbl-6}}
\tablehead{
\colhead{Source ID} & \colhead{R.A.} & \colhead{Decl.} & \colhead{$S_\mathrm{33\,GHz}$} & \colhead{$f_\mathrm{H_{\alpha}}/10^{-13}$} & \colhead{$f_\mathrm{24\,\micron}$} & \colhead{$r_\mathrm{G}$}\\
\colhead{} & \colhead{(J2000)} & \colhead{(J2000)} & \colhead{(mJy)} & \colhead{(erg\,s$^{-1}$\,cm$^{-2}$)} & \colhead{(mJy)} & \colhead{(kpc)}
}
\startdata
                    NGC\,0337\,a   &           $00~59~50.02$  &           $-07~34~33.9$  & $  0.23\pm  0.01$  & $  1.50\pm  0.30$  &$  48.04\pm  2.40$  & 0.945  \\
                    NGC\,0337\,b   &           $00~59~50.68$  &           $-07~34~57.6$  & $  0.62\pm  0.02$  & $  1.72\pm  0.34$  &$  80.48\pm  4.02$  & 1.966  \\
                    NGC\,0337\,c   &           $00~59~51.99$  &           $-07~34~54.9$  & $  0.10\pm  0.01$  & $  0.52\pm  0.10$  &$   8.54\pm  0.43$  & 3.227  \\
                    NGC\,0337\,d   &           $00~59~52.16$  &           $-07~34~38.2$  & $  0.22\pm  0.01$  & $  1.32\pm  0.26$  &$  24.25\pm  1.21$  & 4.061  \\
             NGC\,0628\,Enuc.\,4   &           $01~36~35.72$  &           $+15~50~07.2$  & $  0.12\pm  0.01$  & $  0.49\pm  0.10$  &$  10.30\pm  0.51$  & 7.608  \\
             NGC\,0628\,Enuc.\,2   &           $01~36~37.65$  &           $+15~45~07.2$  & $  0.18\pm  0.02$  & $  0.43\pm  0.09$  &$  42.11\pm  2.11$  & 4.468  \\
             NGC\,0628\,Enuc.\,3   &           $01~36~38.78$  &           $+15~44~23.2$  & $  0.19\pm  0.03$  & $  0.78\pm  0.16$  &$  31.08\pm  1.55$  & 5.720  \\
                       NGC\,0628   &           $01~36~41.70$  &           $+15~46~59.0$  &         $<$  0.06  & $  0.06\pm  0.01$  &$   2.37\pm  0.12$  & 0.071  \\
\enddata
\end{deluxetable*}

\end{document}